\documentclass[11pt]{article}
\pdfoutput=1

\usepackage{amstext,amsmath,amssymb,amsfonts,bbm}
\usepackage{hyperref}
\usepackage{authblk}
\usepackage{caption}
\usepackage[latin1]{inputenc}
\usepackage{epsfig}
\usepackage{amsthm}
\usepackage{subfigure}
\usepackage{color}
\usepackage{graphicx}
\usepackage{slashed}
\usepackage{cite}

\usepackage[lmargin=80pt,rmargin=80pt,tmargin=80pt,bmargin=80pt]{geometry}

\newcommand{\be}{\begin{equation}}
\newcommand{\ee}{\end{equation}}
\newcommand{\nn}{\nonumber}
\newcommand{\f}{\frac}
\newcommand{\p}{\partial}

\newcommand{\Tr}{{\rm Tr}}

\newcommand{\la}{\langle}
\newcommand{\ra}{\rangle}
\newcommand{\tj}[6]{ \begin{pmatrix}
       #1 & #2 & #3 \\
       #4 & #5 & #6 
\end{pmatrix}}
\newcommand{\sj}[6]{ \begin{Bmatrix}
       #1 & #2 & #3 \\
       #4 & #5 & #6 
  \end{Bmatrix}}

\newtheorem*{proposition*}{Proposition}

\theoremstyle{remark}

\DeclareMathOperator{\im}{\mathrm{i}}

\let\a=\alpha   \let\g=\gamma  \let\d=\delta
\let\z=\zeta     \let\th=\theta   \let\l=\lambda
\let\m=\mu    \let\n=\nu           
    \let\vph=\varphi  
    
\let\G=\Gamma \let\D=\Delta    \let\X=F
         
  \let\eps=\epsilon

\newcommand{\Rt}{\tilde{R}}

\newcommand{\phit}{\tilde{\phi}}

\newcommand{\cG}{\mathcal{G}}
\newcommand{\cH}{\mathcal{H}}

\newcommand{\cN}{\mathcal{N}}

\begin{document}

\title{\bf 
$SO(3)$-invariant phase of the $O(N)^3$ tensor model}

\author[1]{Dario Benedetti}
\author[2]{Ilaria Costa}

\affil[1]{\normalsize\it CPHT, CNRS, Ecole Polytechnique, Institut Polytechnique de Paris, \authorcr Route de Saclay, 91128 Palaiseau, France
  \authorcr \hfill }

\affil[2]{\normalsize\it Dipartimento di Fisica Ettore Pancini, Universit\`a di Napoli Federico II, 
Napoli, Italy  \authorcr \hfill}


\date{}

\maketitle

\hrule\bigskip

\begin{abstract}

We study classical and quantum (at large-$N$) field equations of bosonic tensor models with quartic interactions and $O(N)^3$ symmetry.
Among various possible patterns of spontaneous symmetry breaking we highlight an $SO(3)$ invariant solution, with the tensor field expressed in terms of the Wigner $3jm$ symbol.
We argue that such solution has a special role in the large-$N$ limit, as in particular its scaling in $N$ can provide an on-shell justification for the melonic large-$N$ limit of the two-particle irreducible effective action in a broken phase.

\end{abstract}

\hrule\bigskip

\tableofcontents
\newpage

\section{Introduction}

Spontaneous symmetry breaking is an extensively studied subject in statistical and quantum field theory, as it is a paradigm of phase transitions, and it offers the basis for many theoretical developments and phenomenological  models.
Scalar field theories provide the easiest case study, and in particular, the case of $\cN$ scalar fields with an $O(\cN)$-invariant action represents the standard example for introducing Goldstone bosons. A non-zero vacuum expectation value for any of the $\cN$ fields breaks the $O(\cN)$ invariance, leaving only a stabilizer subgroup $O(\cN-1)$ intact, thus giving rise to $\cN-1$ Goldstone bosons that take value in the quotient $O(\cN)/O(\cN-1)$.

Many other patterns of symmetry breaking can be constructed from the same field content if some of the $O(\cN)$ symmetry is explicitly broken by the potential. If the latter is only invariant under a group $\cG \subset O(\cN)$, different non-trivial vacua with different stabilizer subgroups can exist.
We will consider a very concrete example in this paper: we will take $\cN = N^3$, and we will choose a potential with $O(N)^3$ symmetry, which is of course much smaller than $O(N^3)$: the number of generators of $O(N)^3$ is $3N(N-1)/2$, which also acts as an upper bound on the number of Goldstone bosons, well below their number in the $O(N^3)$ case.
A model of this type is naturally interpreted as a rank-3 tensor model, since the $N^3$ fields can be viewed as a tri-fundamental representation of $O(N)^3$, and written as $\phi_{abc}(x)$, with $a,b,c=1,\ldots,N$ and $x\in \mathbb{R}^d$.

Various versions of rank-$r$ tensor models  (by which we mean models with fields transforming in the product of $r\geq 3$ fundamental representations)  have been studied for some years for several reasons, in particular in connection with their large-$N$ limit, which is dominated by melonic diagrams \cite{critical,review,uncoloring,Carrozza:2015adg}. For $d=0$ they can be viewed as models of random $r$-dimensional geometries, or Euclidean quantum gravity \cite{Ambjorn:1990ge,Sasakura:1990fs}, in a similar fashion as matrix models provide a model of two-dimensional quantum gravity \cite{DiFrancesco:1993nw}. For $d=1$, with Grassmann rather than bosonic fields, they provide an alternative to the SYK model, without the need of quenched disorder \cite{Witten:2016iux,Klebanov:2016xxf,Krishnan:2017lra,Choudhury:2017tax}. Lastly, for $d\geq 2$, they represent a novel class of quantum field theories for which we can hope to control non-perturbative aspects via the large-$N$ limit, and in particular discover new interacting conformal field theories \cite{Giombi:2017dtl,Prakash:2017hwq,Benedetti:2017fmp,Giombi:2018qgp,Benedetti:2018ghn,Benedetti:2019eyl,Benedetti:2019ikb,Benedetti:2019rja}.

It is mainly their applications for $d\geq 1$ that motivate our study of spontaneous symmetry breaking (SSB) in tensor models. More precisely, there are two precise question to which we hope to contribute by improving our understanding of possible broken phases in these models.

First, as already mentioned, SSB is typically associated to a phase transition. The latter, in the case of continuous phase transitions, is characterized by a universal critical behavior associated to the properties of a renormalization group fixed point lying on the boundary between two or more phases.
As new classes of fixed points are being identified in tensor models in $d\geq 1$ \cite{Giombi:2017dtl,Prakash:2017hwq,Benedetti:2017fmp,Giombi:2018qgp,Benedetti:2018ghn,Benedetti:2019eyl,Benedetti:2019ikb}, it is natural to ask if they are associated to a phase transition, and in case of what kind. Understanding SSB in tensor models is a first step in order to address that.
Notice that on one hand fixed points of multi-field scalar theories with symmetry group smaller than $O(\cN)$ have been studied to a large extent (see for example Ref.~\cite{Pelissetto:2000ek,Rychkov:2018vya} for reviews), but the ones arising in tensor models are relatively new, and for the moment tightly associated to the large-$N$ limit. On the other hand, SSB in tensor models has been mostly unexplored, with the only exception being the work of Diaz and Rosabal \cite{Diaz:2018eik}, which however, as we will argue in the following, did not provide interesting symmetry breaking solutions from the point of view of the large-$N$ limit.

Our second motivation derives from the development of the two-particle irreducible (2PI) formalism for tensor models. 
Such formalism  is based on a generalization of the 1PI effective action: one starts by defining a free energy in the presence of a bilocal source for the product of two fields as well as the standard source for the fundamental field, and then takes a Legendre transform with respect to both sources, thus defining the 2PI  effective action (see \cite{Cornwall:1974vz}, or the more recent review \cite{Berges:2004yj}). As a consequence, the 2PI effective action is in general a non-local functional depending both on a local field (the one point function), and on a bilocal one (the two-point function). However, in the large-$N$ limit the potential can be greatly reduced: for example, in the case of vector models the potential becomes local \cite{Berges:2004yj}. For the Sachdev-Ye-Kitaev model \cite{Sachdev:1992fk,Kitaev,Polchinski:2016xgd,Maldacena:2016hyu} there exists a bilocal formulation, which can equivalently be derived from the 2PI effective action \cite{Benedetti:2018goh}: in this case, the melonic dominance at large $N$ leads to a bilocal 2PI effective action. Such formulation of the SYK model is very useful, as it allows to: derive the large-$N$ Schwinger-Dyson equations as equations of motion \cite{Kitaev}; neatly identify the light mode associated to the non-conformal perturbation in the strong coupling limit and derive the Schwarzian action controlling its effective dynamics  \cite{Maldacena:2016hyu,Jevicki:2016bwu,Kitaev:2017awl}; efficiently build a perturbative expansion for $n$-point functions of bilinear operators \cite{Jevicki:2016ito,Gross:2017hcz,Gross:2017aos,Kitaev:2017awl}; possibly provide a holographic interpretation along the lines of \cite{Das:2003vw,Koch:2010cy}.
The important role played by a bilocal action formulation for the SYK model, acted as an incentive to the introduction of the 2PI formalism for tensor models in Ref.~\cite{Benedetti:2018goh}, leading to similar results at the first few orders of the $1/N$ expansion. However, the symmetric phase was assumed from the outset in Ref.~\cite{Benedetti:2018goh}, thus restricting to a zero one-point function and a diagonal two-point function.
Here, we wish to make a first step towards extending the 2PI construction to the broken phase, so to build a full effective action depending not only on the two-point function $G$, but also on the one-point function $\phi$. 
We will argue that knowledge of possible non-trivial solutions of the field equations can provide a guiding principle for the large-$N$ expansion of the full 2PI effective action.

The main result of our paper is the finding of an $SO(3)$-invariant solution of the classical field equations of the $O(N)^3$ model with quartic interactions, a model first introduced in zero dimensions in  Ref.~\cite{Carrozza:2015adg} and later in one  and higher dimensions in Ref.~\cite{Klebanov:2016xxf,Giombi:2017dtl,Benedetti:2019eyl}. After discrete Fourier transform, the solution is proportional to a Wigner $3jm$ symbol, and it highlights an intriguing link between tensor models and $SO(3)$ recoupling theory \cite{Yutsis:1962vcy,Varshalovich:1988ye,Brink:1993}.\footnote{Similar solutions of models with tensorial type of interactions have appeared in Ref.~\cite{Sasakura:2005js,Sasakura:2006pq} and \cite{BenGeloun:2018eoe}. However, in both cases the $SO(3)$ (or $SU(2)$) group was built in from the outset, while in our model we start with a much bigger symmetry group, namely $O(N)^3$, and the main point is the appearance of the symmetry reduced solution. At a technical level, as we will see in Sec.~\ref{sec:sol}, we will need to go through a discrete Fourier transform on the tensor indices in order to uncover contraction patterns that are reminiscent (although more general) of $SO(3)$ recoupling theory.} The derivation and analysis of such a solution will be presented in Sec.~\ref{sec:sol}, while in Sec.~\ref{sec:2PI} we will argue that the same type of solution is in principle possible also in the full $2PI$ quantum effective action. The on-shell evaluation of the latter then leads us to dealing with $3nj$ symbols, and on the basis of an old conjecture on their general asymptotics due to Amit and Roginsky \cite{Amit:1979ev}, we conclude that melonic diagrams remain dominant at large $N$.

In App.~\ref{app:3j} we collect useful formulas and notations, while in App.~\ref{app:othersol} we discuss other solutions of the classical field equations, which however are less interesting from the large-$N$ point of view, as they either become trivial or they are only sensitive to one of the three interaction terms.

\section{Spontaneous symmetry breaking at the classical level}
\label{sec:sol}

We consider a real rank-$3$ tensor field, $\phi_{abc}(x)$, with $a,b,c=1,\ldots,N$, 
transforming in the tri-fundamental representation of $O(N)^3$. Notice that no symmetries under permutations of the indices are assumed.
The classical action of the model we consider is:\footnote{We will work in Euclidean signature for simplicity. Repeated indices are summed over. We will in general omit the $x$-dependence of $\phi$, unless fields at different points appear in the same expression.}
\begin{align} \label{eq:action}
S[\phi] &= \int d^d x \f{1}{2}  \phi_{abc} (   - \p_\m\p^\m)^{\zeta} \phi_{abc} + S_{\rm int}[\phi] \, ,\\
 \label{eq:int-action}
S_{\rm int}[\phi] &= \int d^d x  \left(\f12 \l_2 \phi_{abc} \phi_{abc} + \f{\l_d}{4 N^3}  I_d  + \f{\l_p}{4 N^2} I_p+ \f{\l_t}{4 N^{3/2}} I_t \right)\,,
\end{align}
where $\zeta>0$ is a free parameter (but we have in mind either $\zeta=1$, as in Ref.~\cite{Klebanov:2016xxf,Giombi:2017dtl}, or $\zeta=d/4$, as in Ref.~\cite{Benedetti:2019eyl,Benedetti:2019ikb}).
$I_d$, $I_p$ and $I_t$ are the double-trace, pillow, and tetrahedron interactions, respectively:
\begin{align}
I_d & = (\phi_{abc} \phi_{abc})^2 \,,\\
I_p &=  \a_1 \,\phi_{abc} \phi_{ab'c'} \phi_{a'b'c'} \phi_{a'bc} 
+ \a_2\, \phi_{abc}\phi_{a'bc'} \phi_{a'b'c'} \phi_{ab'c}+ \a_3\, \phi_{abc}\phi_{a'b'c} \phi_{a'b'c'} \phi_{abc'} \,,\\
I_t &= \phi_{abc}\phi_{ab'c'} \phi_{a'bc'} \phi_{a'b'c} \,.
\end{align}
We could of course absorb one of the $\a_i$ parameters of $I_p$ in the coupling $\l_p$, but we choose not to do so, in order to keep the freedom to set any of them to zero. 

It is convenient to introduce a graphical representation of the $O(N)^3$ invariants, which also justifies the names of the different contraction patterns. We  represent every tensor as a vertex and every contraction of two indices as an edge. 
We assign to each edge a label (also called \emph{color}) $1$, $2$, or $3$, corresponding to the position of the three indices in the tensor. The resulting graphs are then 3-colored graphs \cite{RTM}. In fact, for our model the labeling will only matter for the pillow interaction, the others being invariant under permutations of the labels, hence we will omit colors, and only add a label when needed.
The interacting part of the action, Eq.~\eqref{eq:int-action}, can then be represented as:
\be \label{eq:int-action-graph}
\begin{split}
S_{\rm int}[\phi] = & \int d^d x  \left(\f12 \l_2 \,\vcenter{\hbox{\includegraphics[width=1.6cm]{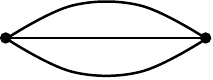}}}
+ \f{\l_d}{4 N^3} \, \vcenter{\hbox{\includegraphics[width=1.6cm]{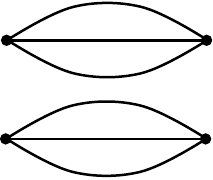}}} \right.\\
&\qquad\left.  + \f{\l_p}{4 N^2}  \sum_{i=1,2,3} \a_i \,\vcenter{\hbox{\includegraphics[width=1.6cm]{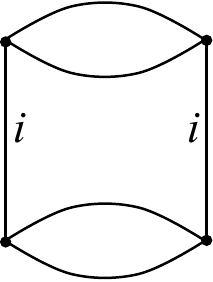}}}  
 +\f{\l_t}{4 N^{3/2}} \,\vcenter{\hbox{\includegraphics[width=1.6cm]{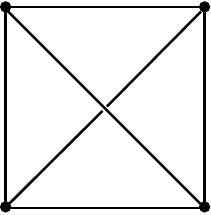}}} \right)\,.
\end{split}
\ee

The scaling in $N$ of the couplings is chosen in such a way that the Feynman diagrams of the theory have a non-trivial large-$N$ limit (in the sense that it exists and that it contains an infinite family of graphs) \cite{Carrozza:2015adg}.
As $I_d$ and $I_p$ are positive definite (for positive $\a_i$ at least, as we assume), we will assume from now on that $\l_d$ and $\l_p$ are positive.
On the contrary, the invariant $I_t$ has indefinite sign and is unbounded, therefore the sign of its coupling cannot be fixed by stability requirements.
In fact it could even be taken purely imaginary \cite{Benedetti:2019eyl,Benedetti:2019ikb}, similarly to the $\varphi^3$ interaction in the Lee-Yang model \cite{Fisher:1978pf,Cardy:1985yy}. In this paper we will mostly assume $\l_t\in \mathbb{R}$, because we are interested in real solutions of the equations of motion, and we will assume that the theory makes sense in the large-$N$ limit.\footnote{In matrix models unstable potentials are very common, with large-$N$ critical points typically located well within the unstable region \cite{DiFrancesco:1993nw}. In tensor models in zero dimension a similar situation is very common \cite{critical,uncoloring}.}

The action is invariant under two types of symmetries: spacetime and internal symmetries. The first correspond to invariance under the Euclidean group of transformations $ISO(d)$.
The second are given by  $O(N)^{3}$ transformations (in the tri-fundamental representation):\footnote{It is important to stress the difference between the tri-fundamental representation of $O(N)^3$ and the representation of $O(N)$ constructed as a product of fundamentals. In the former case the three rotation matrices in Eq.~\eqref{eq:ON3} are in general different, while in the latter they would be the same. Tensor models with $O(N)$ symmetry are more complicated to analyze, and they have a well-defined large-$N$ limit only if one restricts to one of the irreducible components of the product of representations, as  shown in Ref.~\cite{Benedetti:2017qxl,Carrozza:2018ewt} following a conjecture made in Ref.~\cite{Klebanov:2017nlk}.}
\be \label{eq:ON3}
\phi_{abc} \to \phi'_{abc} = R^{(1)}_{aa'} R^{(2)}_{bb'} R^{(3)}_{cc'} \phi_{a'b'c'}\,, \;\;\;\; R^{(i)}\in O(N)\,.
\ee
In addition, for $\a_1=\a_2=\a_3$ the model is also invariant under permutations of the indices, a symmetry known in the literature as  color symmetry, because it is equivalent to a permutation of the colors in the 3-colored graphs. The interacting action we wrote is the most general derivative-free action with such symmetries and at most quartic in the fields.

The field equations are obtained by imposing that the first functional variation with respect to $\phi_{abc}$ is zero:
\be \label{eq:eom}
\begin{split}
0= &\left((-\p_\m\p^\m)^\z   + \l_2 + \f{\l_d}{ N^3}  (\phi_{a'b'c'} \phi_{a'b'c'})\right) \phi_{abc} \\
&+ \f{\l_p }{N^2} (\a_1\, \phi_{ab'c'} \phi_{a'b'c'} \phi_{a'bc} +\a_2\, \phi_{a'bc'} \phi_{a'b'c'} \phi_{ab'c}+\a_3\, \phi_{a'b'c} \phi_{a'b'c'} \phi_{abc'})\\
& +  \f{\l_t}{N^{3/2}} \phi_{ab'c'} \phi_{a'bc'} \phi_{a'b'c} \,.
\end{split}
\ee

We will only consider solutions that do not break Euclidean invariance, that is, we will consider the case of constant field configurations, for which the Laplacian on the first line drops out. In this respect, the spacetime dimension plays no role, and we could have taken $d=0$ from the beginning. However, the spacetime dimension plays a role at the quantum or statistical level in determining whether spontaneous symmetry breaking actually occurs: by the Coleman-Mermin-Wagner theorem \cite{Mermin:1966fe,Coleman:1973ci} we do not expect spontaneous symmetry breaking of continuous symmetries for $d\leq 2$ (for $\z=1$).

Discarding the derivative term, we have the following graphical representation of the field equations:
\be
\begin{split}
0 =   \l_2 \,\vcenter{\hbox{\includegraphics[width=.7cm]{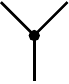}}}
+ \f{\l_d}{N^3} \, \vcenter{\hbox{\includegraphics[width=1.9cm]{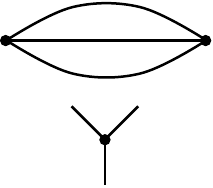}}} 
+ \f{\l_p}{N^2}  \sum_{i=1,2,3} \a_i \,\vcenter{\hbox{\includegraphics[width=.7cm]{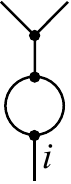}}}  
 +\f{\l_t}{N^{3/2}} \,\vcenter{\hbox{\includegraphics[width=1.6cm]{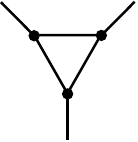}}} \,,
\end{split}
\ee
where the open half-edges represent the free indices.

An obvious solution of the field equations is $\phi_{abc} =0$, which of course is compatible with all the symmetries of the model.
However, we are interested in studying non-zero solutions, which generally are possible when $\l_2<0$.

Any non-zero solutions will necessarily break the $O(N)^{3}$ symmetry, because the three copies of $O(N)$ act independently on each index, and because there exists no non-trivial vector which is invariant under the action of $O(N)$.
How much of the symmetry group will be broken depends on the specific solution.
For an $N$-component vector, the subgroup of $O(N)$ that leaves the vector invariant (i.e.\ the stabilizer) is clearly $O(N-1)$ because it corresponds to rotations in the hyperplane orthogonal to the vector. However, for tensors $O(N-1)^{3}$ is not in general the stabilizer of a solution, and there exist in principle many possible patterns of symmetry breaking.

We would like to identify solutions such that the on-shell action scales homogeneously in $N$. Such solutions must have a non-trivial scaling in $N$, because all the quartic invariants have the same power of the field and the same number of summations, but different powers of $N$ in front.
In particular, for such a solution, each term in the field equations \eqref{eq:eom} should be proportional to $\phi_{abc}$, with proportionality factor of order $N^0$, as in the mass term. If any term would have a scaling as $N^\a$, with $\a\neq 0$, it would either not contribute in the large-$N$ limit ($\a<0$), or dominate and lead to a trivial equation ($\a>0$).
Therefore, the solutions with good large-$N$ behavior should satisfy:
\begin{align} \label{eq:phi-eq-1}
\phi_{a'b'c'} \phi_{a'b'c'} & \sim N^3 \,, \\
 \label{eq:phi-eq-2}
\phi_{ab'c'} \phi_{a'b'c'} & \sim N^2 \d_{aa'} \,, \\
 \label{eq:phi-eq-3}
\phi_{ab'c'} \phi_{a'bc'} \phi_{a'b'c} & \sim N^{3/2} \phi_{abc} \,,
\end{align}
plus two other equations similar to the second above, but with free indices in second or third position, if $\a_i\neq 0$, $\forall i$.

We will now present a solution with the above properties, while in App.~\ref{app:othersol} we will discuss solutions that do not have a good scaling in $N$, as well as others which do, but exist only for special restrictions of our model (with some couplings set to zero).

\subsection{ $SO(3)$-invariant solution}
\label{sec:Fourier}

We will restrict from now on to odd $N$, and we will set $N=2j+1$.
In order to uncover a non-trivial solution of the  field equations \eqref{eq:eom}, it is convenient to perform a discrete Fourier transform in index space.
More precisely, let\footnote{In Fourier space we do not use Einstein's convention, i.e.\ repeated indices are not automatically summed over.}
\be \label{eq:phiFourier}
\phi_{abc} = \sum_{m_1=-j}^j\sum_{m_2=-j}^j\sum_{m_3=-j}^j\phit_{m_1 m_2 m_3} \im^{3j-|m_1|-|m_2|-|m_3|} e^{-\f{2\pi \im}{N} (a m_1+b m_2 + c m_3)} \,,
\ee
where we have rescaled the Fourier transformed field by a factor $\im^{3j-|m_1|-|m_2|-|m_3|}$ for later convenience.
Similarly we can transform the rotation matrices $R\in O(N)$:
\be \label{eq:R-Fourier}
R_{ab} = \sum_{m_1=-j}^j\sum_{m_2=-j}^j \Rt_{m_1 m_2} \im^{2j-|m_1|-|m_2|} e^{-\f{2\pi \im}{N} (a m_1+b m_2)} \,.
\ee
It is also convenient to introduce the matrix
\be \label{eq:metric}
g_{m m'} = g^{m m'} \equiv  (-1)^{j-m} \d_{m\, -m'} \,,
\ee
which can be recognized to be the $SU(2)$-invariant metric; unlike in App.~\ref{app:3j}, Eq.~\eqref{eq:metric_j}, we omit here the subscript $j$, as we do not need to mix different $j$'s.
We will use the matrix \eqref{eq:metric} to raise or lower indices (notice that $\sum_{m''} g_{m m''} g^{m'' m'} = \d_m^{m'}$).

The reality of $\phi_{abc}$ and $R_{ab}$ implies the reality conditions
\be
\label{eq:reality-cond}
\phit^*_{m_1 m_2 m_3}  = \phit_{-m_1 -m_2 -m_3} (-1)^{3j-|m_1|-|m_2|-|m_3|} \equiv \phit^{m_1 m_2 m_3} \,,
\ee
\be
\Rt^*_{m_1 m_2}  = \Rt_{-m_1 -m_2} (-1)^{2j-|m_1|-|m_2|} \equiv \Rt^{m_1 m_2}  \,.  \label{eq:R-reality-cond}
\ee

Using the identity
\be \label{eq:delta_m}
\sum_{a=1,\ldots,N} e^{-\f{2\pi \im}{N} a (m + m')} = N \d_{m\,-m'} \,,
\ee
it is easily checked that the transform \eqref{eq:phiFourier} has two effects on the action \eqref{eq:int-action}:
\begin{enumerate}
\item it trades every contraction of two indices in direct space for a contraction of two indices in Fourier space, but with a relative sign in the latter case (i.e.\ an index $m$ is contracted with an index $-m$);
\item for every sum over an index $m$, it introduces a multiplicative factor $(-1)^{j-|m|}=(-1)^{j-m}$, where we use the fact that $j$, and hence $m$, is an integer.
\end{enumerate}
In other words, index contractions  in Fourier space are done by means of the metric \eqref{eq:metric}.
Although we will not need it, the relative signs in the contraction of two indices can be accounted for in the graphical representation  by the addition of an arrow on every edge, with the following convention: an arrow directed away from (towards) a vertex indicates a positive (negative) sign of the corresponding index $m$ of the tensor represented by the vertex.

Furthermore, the transformed rotation matrices satisfy:
\be \label{eq:Fourier-orthog}
 \sum_m \Rt_{m_1 m} \Rt_{m_2}{}^m = \f{g^{m_1 m_2}}{N^2}\,,
\ee
and the transformed action is invariant under
\be \label{eq:Fourier-symm}
\phit_{m_1 m_2 m_3} \to \phit'_{m_1 m_2 m_3} = \sum_{m'_1, m'_2, m'_3} \Rt^{(1)}_{m_1}{}^{m'_1} \Rt^{(2)}_{m_2}{}^{m'_2} \Rt^{(3)}_{m_3}{}^{m'_3} \, \phit_{m'_1 m'_2 m'_3} \,.
\ee

The new contraction rules, as well as the graphical representation, will probably ring a bell to the reader familiar with the theory of angular momentum, and in particular with the composition of Wigner coefficients (e.g. \cite{Yutsis:1962vcy,Varshalovich:1988ye,Brink:1993}).\footnote{In the theory of angular momentum, graphs have also an assignment of cyclic order at each vertex, but at this stage we do not need to introduce orientations, as the tensor has no symmetries under permutations of the indices, and the $O(N)^3$ symmetry implies that a first index can only be contracted with a first index, etc. Correspondingly, we have instead color labels for the edges.}
We are in fact ready to prove the following:

 \begin{proposition*}
The field equations \eqref{eq:eom} admit, for $N=2j+1$ and $j\in \mathbb{N}$, a solution in the form of Eq.~\eqref{eq:phiFourier}, with the transformed field given by:
\be \label{eq:sol-Fourier}
\phit_{m_1 m_2 m_3} = \Phi_{m_1 m_2 m_3} \equiv  \g\, \tj{j}{j}{j}{m_1}{m_2}{m_3} \,,
\ee
and with
\be \label{eq:sol-c}
\g = \pm \sqrt{ - \f{\l_2}{ \left(\l_d +\l_p (\a_1+\a_2+\a_3) +\l_t N^{3/2} \begin{Bmatrix}
   j & j & j \\
   j & j & j 
  \end{Bmatrix} \right) } } \,,
\ee
where $\tj{j}{j}{j}{m_1}{m_2}{m_3}$ and $\begin{Bmatrix}
   j & j & j \\
   j & j & j 
  \end{Bmatrix} $ are the Wigner $3jm$ and $6j$ symbols, respectively.
\end{proposition*}

\

\begin{proof}

By the transformation \eqref{eq:phiFourier}, the field equations  \eqref{eq:eom}  become:
\be \label{eq:eom-Fourier}
\begin{split}
0= &  \f{1}{N^3} \f{\d S}{\d \phit^{m_1 m_2 m_3}}\\
= &(-\p_\m\p^\m   + \l_2 ) \phit_{m_1 m_2 m_3} \\ 
&+ \sum_{m'_1, m'_2, m'_3} 
\left(\l_d  (\phit_{m'_1 m'_2 m'_3} \phit^{m'_1 m'_2 m'_3})) \phit_{m_1 m_2 m_3} \right.\\
&+ \l_p N (\a_1\, \phit_{m_1 m'_2 m'_3} \phit^{m'_1 m'_2 m'_3} \phit_{m'_1 m_2 m_3} 
+\a_2\, \phit_{m'_1 m_2 m'_3} \phit^{m'_1 m'_2 m'_3} \phit_{m_1 m'_2 m_3}\\
& \quad\qquad +\a_3\, \phit_{m'_1 m'_2 m_3} \phit^{m'_1 m'_2 m'_3} \phit_{m_1 m_2 m'_3})\\
& \left. +\,  \l_t N^{3/2} \phit_{m_1 m'_2}{}^{m'_3} \phit^{m'_1}{}_{m_2 m'_3} \phit_{m'_1}{}^{m'_2}{}_{ m_3} \right) \,.
\end{split}
\ee

In order to check that Eq.~\eqref{eq:sol-Fourier} solves the field equations, we need to use two identities (e.g. \cite{Yutsis:1962vcy,Varshalovich:1988ye,Brink:1993}). The first is a special case of the orthogonality relation of $3jm$-symbols, Eq.~\eqref{eq:orthog1}:
\be \label{eq:3j-orthog}
 \sum_{m'_2, m'_3} (-1)^{2j-m'_2 -m'_3} \tj{j}{j}{j}{m_1}{m'_2}{m'_3} \tj{j}{j}{j}{-m'_1}{-m'_2}{-m'_3} =  \f{1}{N}  g^{m_1 m'_1}\,,
\ee
which translates into
\be \label{eq:phi-orthog}
\begin{split}
\sum_{m'_2, m'_3} &  \Phi_{m_1 m'_2 m'_3} \Phi^{m'_1 m'_2 m'_3} = \f{\g^2}{N}  \d_{m_1}^{m'_1} \,,
\end{split}
\ee
and therefore, also
\be \label{eq:phi-norm}
\sum_{m'_1,m'_2, m'_3} \Phi_{m'_1 m'_2 m'_3} \Phi^{m'_1 m'_2 m'_3} = \g^2 \,.
\ee
The second is a special case of the identity \eqref{eq:3j-triangle}. 
Adapted to our case with equal $j$'s, and using the invariance of the $3jm$ symbol under even permutations of the columns, we obtain:
\be \label{eq:phi-triangle}
\begin{split}
\sum_{m'_1, m'_2, m'_3} &  \Phi_{m_1 m'_2}{}^{m'_3} \Phi^{m'_1}{}_{m_2 m'_3} \Phi_{m'_1}{}^{m'_2}{}_{ m_3}
= \g^2 \,\begin{Bmatrix}
   j & j & j \\
   j & j & j 
  \end{Bmatrix} \Phi_{m_1 m_2 m_3}  \, .
\end{split}
\ee
Notice that the relative minus signs of the contracted indices in Eq.~\eqref{eq:3j-triangle} is the crucial reason why we need to go through a Fourier transform before using such formula.\footnote{In fact, for $\l_t=0$, the field equations admit the following solution in direct space:
\be \label{eq:3j-sol}
\phi_{abc} = \g_0 N^{3/2}   \tj{j}{j}{j}{a-\f{N+1}{2}}{b-\f{N+1}{2}}{c-\f{N+1}{2}} \,,
\ee
with $\g_0$ equal to $\g$ at $\l_t=0$. The magnetic indices of the $3jm$ symbol have been translated in order to take into account the fact that our tensor indices are defined to take values between 1 and $N$, rather than between $-j$ and $j$.
Although we couldn't find a mapping between the two, and therefore we believe they are distinct solutions, the properties (such as stability, large-$N$, and so on) of this solution are the same as those of Eq.~\eqref{eq:sol-Fourier} at $\l_t=0$.
}

Using Eq.~\eqref{eq:phi-orthog},  \eqref{eq:phi-norm}, and  \eqref{eq:phi-triangle}, we find that all the terms in Eq.~\eqref{eq:eom-Fourier} are proportional to $ \Phi_{m_1 m_2 m_3}$, hence the field equations reduce to an equation for $\g$.
For constant $\g$, this reads:
\be \label{eq:c-eq}
0= \l_2+ \left(\l_d +\l_p (\a_1+\a_2+\a_3) +\l_t N^{3/2} \begin{Bmatrix}
   j & j & j \\
   j & j & j 
  \end{Bmatrix} \right) \g^2 \,,
\ee
and its solution is Eq.~\eqref{eq:sol-c}.

\end{proof}

Notice that for even $N$ (half-integer $j$) we have no solutions of this sort, because the $3jm$ symbol with three identical $j$'s is identically zero in such case. We now proceed to discuss some properties of the solution.

 \paragraph{Large-$N$.}

The equations  (\ref{eq:phi-orthog}-\ref{eq:phi-triangle}), when rewritten for the original field $\phi_{abc}$, take precisely the form of Eq.~(\ref{eq:phi-eq-1}-\ref{eq:phi-eq-3}).
The first two are straightforward. The last one holds in the large-$N$ limit. In fact, the asymptotic behavior of the $6j$ symbol is (e.g. \cite{Bonzom:2008xd})
\be
\begin{Bmatrix}
   j & j & j \\
   j & j & j 
  \end{Bmatrix} = \f{2^{5/4}}{\sqrt{\pi N^3}} \cos\left(3N \arccos\left(-\f{1}{3}\right)+\f{\pi}{4}\right) \left(1+O\left(\f{1}{N}\right)\right) \,,
\ee
with a $1/N^{3/2}$ behavior that exactly compensates the $N^{3/2}$ in the equation \eqref{eq:eom-Fourier}.
Furthermore, the cosine is zero only for
\be
N = \pi \f{4  n+1 }{12 \arccos\left(-\f{1}{3}\right)} \,,
\ee
with integer $n$, but since in such case $N$ is not an integer, we conclude that $\l_t$ always contributes in such limit.
Hence, all the terms in the field equations are of the same order in $N$, as desired. Similarly they all contribute at the same order to the on-shell action, which is of order $N^3$.

 \paragraph{Reality of the solution and sign of the on-shell action.}
 
It is easy to check that the reality condition \eqref{eq:reality-cond} is satisfied by the solution  \eqref{eq:sol-Fourier} for real $\g$.
First, notice that since the $3jm$ symbol is non-zero only for $m_1+m_2+m_3=0$, we have
\be \label{eq:-sum_m}
(-1)^{|m_1|+|m_2|+|m_3|}=(-1)^{m_1+m_2+m_3} = 1\,.
\ee
Next, we use the time-reversal property of $3jm$ symbols, Eq.~\eqref{eq:time-rev}.
Since furthermore the $3jm$ symbols are real, we straightforwardly find that Eq.~\eqref{eq:reality-cond}  is equivalent to $\g=\g^*$.
Hence the solution is real for
\be
\frac{\l_2}{ \left(\l_d +\l_p (\a_1+\a_2+\a_3) +\l_t N^{3/2} \sj{j}{j}{j}{j}{j}{j} \right) }  <0 \,.
\ee
However, the on-shell action will be negative (and thus below the value at the trivial solution) only when the denominator is positive and $\l_2<0$.
To that end, we need $|\l_t|<\pi^{1/2}2^{-5/4}(\l_d +\l_p (\a_1+\a_2+\a_3))$.

 \paragraph{Symmetry of the solution.}

As recalled in App.~\ref{app:3j}, Eq.~\eqref{eq:3j-invariance}, the $3jm$ symbol with three identical $j$ is an invariant tensor under $SO(3)$ rotations in the spin-$j$ representation. It is thus clear that $SO(3)$ is the stability group of our solution $\phit_{m_1 m_2 m_3} = \Phi_{m_1 m_2 m_3}$.
In direct space, such invariance becomes
\be
\phi_{abc}  = R_{aa'} R_{bb'} R_{cc'} \phi_{a'b'c'}\,, 
\ee
with
\be \label{eq:D-Fourier}
\Rt_{m_1}{}^{m_2} = \f{1}{N}  D^{j}_{m_1 m_2}(g) \,,
\ee
and $g\in SO(3)$. Notice that the properties \eqref{eq:Dconj} and \eqref{eq:g-invar} of the Wigner $D$-matrices imply that $\Rt_{m_1 m_2}$ satisfies the reality condition \eqref{eq:R-reality-cond} and the orthogonality relation \eqref{eq:Fourier-orthog}. Hence equations \eqref{eq:R-Fourier} and \eqref{eq:D-Fourier} define an $N$-dimensional orthogonal representation of $SO(3)$.\footnote{Such non-standard representation of $SO(3)$ (or in fact of $SU(2)$) is close to a general class of nonstandard representations introduced in Ref.~\cite{Kibler:1997mb,Kibler:2009bnw,Kibler:2010dzq}.}
Therefore, we expect to find $\f32 N (N-1)-3$ Goldstone modes.

Notice also that the constraint $m_1+m_2+m_3=0$ implies that 
\be
\Phi_{abc} = \Phi_{a+k \, b+k\, c+k} \,,
\ee
that is, the solution is a tensor invariant under simultaneous translation of all its indices.

 \paragraph{Stability.}
 One can verify the stability of the solution by studying the Hessian.\footnote{For $\l_t\neq 0$, at best we expect a local stability, as we know that the tetrahedron interaction is unbounded from below. At large $N$, local stability is typically enough, as tunneling towards the unstable region is exponentially suppressed in $N$ (e.g.\ \cite{DiFrancesco:1993nw}).} The generic form of the Hessian  $\cH$ in direct space is given in App.~\ref{app:othersol}.
 Similarly to the field transform \eqref{eq:phiFourier}, we can transform it in the following way:
 \be
 \cH_{abc,a'b'c'}=\sum_{\{m_i,m'_i\}}
 \tilde \cH_{m_1m_2m_3,m'_1m'_2m'_3} \im^{6j-\sum_i |m_i|-\sum_i |m'_i|} e^{-\f{2\pi \im}{N} (a m_1+b m_2 + c m_3+a'm'_1+b'm'_2+c'm'_3)}\,,
 \ee
which is equivalent to
\be
\tilde\cH_{m_1m_2m_3,m'_1m'_2m'_3}(x)= \f{1}{N^6} \int_{x'}\f{\d^2S_{\rm int}}{\d \phit^{m_1 m_2 m_3}(x)\d\phit^{m'_1 m'_2 m'_3}(x')} \,.
\ee
Evaluating it on shell, we find: 
 \be
 \begin{split}
 	\tilde \cH_{m_1m_2m_3,m'_1m'_2m'_3}=& -\frac{\g^2 \l_t }{N^{3/2}} \begin{Bmatrix}
 			j & j & j \\
 			j & j & j 
 	\end{Bmatrix} g_{m_1m'_1}g_{m_2m'_2}g_{m_3m'_3} \\
			&
 	+\frac{2\l_d}{N^3}\Phi_{m_1m_2m_3}\Phi_{m'_1m'_2m'_3} \\&
	+\frac{\l_p}{N^2}\big(\a_1\Phi_{m'_1m_2m_3}\Phi_{m_1m'_2m'_3}+\a_2\Phi_{m_1m'_2m_3}\Phi_{m'_1m_2m'_3}\\&
	\quad +\a_3\Phi_{m_1m_2m'_3}\Phi_{m'_1m'_2m_3}\big)\\&
 	+\frac{\l_p}{N^2}\Big(\a_1\sum_{m''_1}\Phi_{m''_1m_2m_3}\Phi^{m''_1}{}_{m'_2m'_3}g_{m_1m'_1}\\&
 	\quad+\a_2\sum_{m''_2}\Phi_{m_1m''_2m_3}\Phi_{m'_1}{}^{m''_2}{}_{m'_3}g_{m_2m'_2}\\&
 	\quad+\a_3\sum_{m''_3}\Phi_{m_1m_2m''_3}\Phi_{m'_1m'_2}{}^{m''_3}g_{m_3m'_3}\Big)\\&
 	+\frac{\l_t}{N^{3/2}}\Big(\sum_{m''_1}  \Phi_{m''_1m'_2m_3}\Phi^{m''_1}{}_{m_2m'_3}g_{m_1m'_1}\\&
	\quad+\sum_{m''_2}\Phi_{m'_1m''_2m_3}\Phi_{m_1}{}^{m''_2}{}_{m'_3}g_{m_2m'_2}+\sum_{m''_3}\Phi_{m_1m'_2m''_3}\Phi_{m'_1m_2}{}^{m''_3}g_{m_3m'_3}\Big)\,.
 \end{split}
 \ee
In graphical representation we have
\be
\begin{split}
\f{1}{\g^2 } \tilde \cH_{m_1m_2m_3,m'_1m'_2m'_3}  = & \,-\frac{\l_t }{N^{3/2}} 
 \vcenter{\hbox{\includegraphics[width=1.1cm]{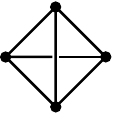}}}
  \,\vcenter{\hbox{\includegraphics[width=1.2cm]{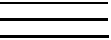}}}
+ \f{2 \l_d}{N^3} \,  \vcenter{\hbox{\includegraphics[width=1.2cm]{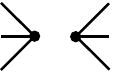}}}  \\
& + \f{\l_p}{N^2}  \sum_{i=1,2,3} \a_i \,\left(\vcenter{\hbox{\includegraphics[width=.8cm]{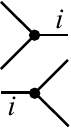}}}  +  \vcenter{\hbox{\includegraphics[width=1.2cm]{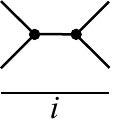}}}  \right) \\
& +\f{\l_t}{N^{3/2}} \sum_{i=1,2,3} \, \vcenter{\hbox{\includegraphics[width=1.2cm]{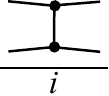}}} \,,
\end{split}
\ee
where now the vertices represent $3jm$ symbols, and horizontal lines without vertices represent the invariant metrics.
The graphical representation here is intended only as a rough guidance: in order to be used for actual computations, one should be careful about labels, and about the orientation of the vertices, which matters once we go on shell, due to the symmetry properties of the $3jm$ symbol.

For the stability analysis we expand the field, $\phit_{m_1 m_2 m_3}  = \Phi_{m_1 m_2 m_3} +\d\phit_{m_1 m_2 m_3}$, and we need to establish the positive definiteness of the quadratic form
\be
\begin{split}
 \sum_{\{m_i,m'_i\}}   \d\phit_{m_1 m_2 m_3}  \tilde \cH^{m_1m_2m_3}{}_{m'_1m'_2m'_3} \d\phit^{m'_1 m'_2 m'_3}\,.
\end{split}
\ee
We thus need to solve the eigenvalue problem
\be \label{eq:eig-H}
\sum_{\{m_i'\}}\tilde \cH^{m_1m_2m_3}{}_{m'_1m'_2m'_3} \chi^{m'_1 m'_2 m'_3} = \theta \chi^{m_1 m_2 m_3} \,,
\ee
with the reality constraint
\be \label{eq:eig-reality}
\chi_{m_1 m_2 m_3}^* = \chi^{m_1 m_2 m_3} \,,
\ee
so that writing $ \d\phit^{m_1 m_2 m_3}= \sum_\n c_\n \chi^{(\n)}{}^{m_1 m_2 m_3}$, with real coefficients $c_\n$, the quadratic form reduces to
\be
\label{quadform}
\sum_\n \theta_\n c_\n^2 \sum_{\{m_i\}}\chi^{(\n)}_{m_1 m_2 m_3} \chi^{(\n)}{}^{m_1 m_2 m_3} = 
\sum_\n \theta_\n c_\n^2 \sum_{\{m_i\}}\chi^{(\n)}_{m_1 m_2 m_3} \chi^{(\n)\,*}_{m_1 m_2 m_3}\,.
\ee
The solution is linearly stable if and only if $\theta_\n\geq 0$. The zero eigenvalues correspond to either exactly flat directions, such as those associated to Goldstone modes, or to directions that are flat only at linear order and which might be either stable or unstable beyond the linear order.

We will sometimes use the condensed notation $\cH_{\{m\},\{m'\}}\equiv  \cH_{m_1m_2m_3,m'_1m'_2m'_3}$ whenever we want to emphasize that we think of a triplet of indices as of a single label for a vector or matrix component. The matrix $\cH_{\{m\},\{m'\}}$ is real and symmetric, and it commutes with the matrix $\D_{\{m\},\{m'\}}  \equiv g_{m_1m'_1}g_{m_2m'_2}g_{m_3m'_3}$ (thanks to properties \eqref{eq:m0} and \eqref{eq:time-rev}). Therefore, it has real eigenvalues, and its (real) eigenvectors can be chosen to be also eigenvectors of $\D_{\{m\},\{m'\}}$. Since the latter has eigenvalues $\pm 1$, the eigenvectors can be divided in two sets, a ``plus'' set corresponding to the $+1$ eigenvalues  of $\D_{\{m\},\{m'\}}$, and a ``minus'' set corresponding to the $-1$ eigenvalues. The eigenvectors satisfying Eq.~\eqref{eq:eig-H} and the reality constraint \eqref{eq:eig-reality} are given by the common eigenvectors of $\cH_{\{m\},\{m'\}}$ and $\D_{\{m\},\{m'\}}$, with the eigenvectors in the minus set being multiplied by $\im$.

The exact diagonalization in the general case turns out to be too complicated, but we will be able to perform it completely for $\l_t=0$, and achieve a block-diagonal form otherwise.

In order to construct the eigenvectors of the Hessian, we begin by defining a basis for rank-3 tensors in terms of $3jm$ symbols, or rather $4jm$ symbols. That is, we define:
\be \label{eq:e-basis}
\begin{split}
e_{m_1m_2 m_3}(L,J,\m) &\equiv \sum_{k} (-1)^{L-k}  \tj{j}{j}{L}{m_1}{m_2}{k} \tj{L}{J}{j}{-k}{\m}{m_3} \\ 
& \equiv \begin{pmatrix}
       j & j & j & J \\
       m_1 & m_2 & m_3 & \m 
\end{pmatrix}^{((1+2)+3)}_L\,,
\end{split}
\ee
where in the last line we have adopted the notation of  Ref.~\cite{Yutsis:1962vcy} in order to point out that such objects are a particular generalized Wigner coefficient, corresponding to the addition of three angular momenta according to the rule $((1+2)+3)$.
Here, $L$, $J$ and $\m$ are treated as free parameters, labelling the different eigenvectors. They must satisfy the relevant triangular inequalities, hence they take values in the ranges
\be
L=0,\ldots,2j\,, \;\;\; J=|j-L|,\ldots,j+L\,,\;\;\; \m=-J,\ldots,+J\,,
\ee
or, equivalently,
\be
J=0,\ldots,3j\,, \;\;\; L=|j-J|,\ldots,{\rm min}\{2j,j+J\}\,,\;\;\; \m=-J,\ldots,+J\,.
\ee
As a consequence, we have precisely $(2j+1)^3=N^3$ eigenvectors. 
Notice that the vectors $e_{\{m\}}(L,J,\m)$ are orthogonal, but not normalized, i.e.\ they satisfy 
\be
\sum_{\{m\}} e_{m_1m_2 m_3}(L,J,\m)e_{m_1m_2 m_3}(L',J',\m')= \f{1}{(2J+1)(2L+1)} \d_{LL'}\d_{JJ'}\d_{\m\m'}\,.
\ee
By direct inspection, we find that on such basis the Hessian is block diagonal, as the labels $J$ and $\m$ are unaffected by its action.
The reality condition is also easily taken into account, because of the following identity
\be
e^{m_1m_2 m_3}(L,J,\m) = (-1)^{J-\m} e_{m_1m_2 m_3}(L,J,-\m) \;.
\ee
We introduce the new basis
\be
e^\eta_{m_1m_2 m_3}(L,J,\m) = \sqrt{\f{\eta(-1)^{J-\m}}{2}(2L+1)(2J+1)}\left( e_{m_1m_2 m_3}(K,J,\m) +\eta\, e_{m_1m_2 m_3}(K,J,-\m) \right) \;,
\ee
where $\eta=\pm1$, such that it is orthonormal, and satisfies the reality condition \eqref{eq:eig-reality}.

In order to further diagonalize the blocks, we seek a basis of eigenvectors in the form
\be \label{eq:Eig-general}
E^{\eta}_{m_1m_2 m_3}(L,J,\m) = \sum_{K} \a_{{}_{KLJ}}\, e^\eta_{m_1m_2 m_3}(K,J,\m) \;.
\ee
For $\l_t=0$, we find only three sets of eigenvectors that for even $J$ have non-zero, and positive, eigenvalues. The first is:\footnote{Notice that this can be written in the form \eqref{eq:Eig-general}, because 
\be
e^\eta_{m_2 m_3 m_1}(j,J,\m) = \eta(-1)^{J-\m+j} N \sum_{K} \sj{j}{j}{K}{J}{j}{j}\, e^\eta_{m_1m_2 m_3}(K,J,\m) \;,
\ee
and similar for the other permutation of the indices.
}
\be
E^{\eta,1}_{m_1m_2 m_3}(j,J,\m) = e^\eta_{m_1m_2 m_3}(j,J,\m) + e^\eta_{m_2 m_3 m_1}(j,J,\m) + e^\eta_{m_3 m_1 m_2}(j,J,\m) \;,
\ee
with  $0\leq J\leq 2j$ for $j>1$, and $0\leq J\leq 1$ for $j=1$ (the latter being special because $E^{\eta,1}_{m_1m_2 m_3}(1,2,\m)$ is identically zero). The corresponding eigenvalues are
\be \label{eq:theta1}
\theta_1(J) = \f{2 \l_d}{N^3} \d_{0J} + \f{\l_p}{N^2} (1+(-1)^J) \left(\f1N +2 (-1)^j \sj{j}{j}{j}{j}{j}{J} \right) \;,
\ee
of multiplicity $2J+1$. The second and third sets of eigenvectors are
\be
E^{\eta,2}_{m_1m_2 m_3}(j,J,\m) = e^\eta_{m_1m_2 m_3}(j,J,\m) -2 e^\eta_{m_2 m_3 m_1}(j,J,\m) + e^\eta_{m_3 m_1 m_2}(j,J,\m) \;,
\ee
and
\be
E^{\eta,3}_{m_1m_2 m_3}(j,J,\m) = e^\eta_{m_1m_2 m_3}(j,J,\m) + e^\eta_{m_2 m_3 m_1}(j,J,\m) -2 e^\eta_{m_3 m_1 m_2}(j,J,\m) \;,
\ee
with $1\leq J\leq 2j$ for all $j$, and with equal eigenvalues
\be \label{eq:theta2-3}
\theta_{2}(J) = \theta_{3}(J) =  \f{\l_p}{N^2} (1+(-1)^J) \left(\f1N - (-1)^j \sj{j}{j}{j}{j}{j}{J} \right) \;,
\ee
of multiplicity  $2J+1$ each.  

Notice that $E^{\eta,1}_{m_1m_2 m_3}(j,J,\m)$ is a completely symmetric or antisymmetric tensor for $j$ even or odd, respectively, while $E^{\eta,2}_{m_1m_2 m_3}(j,J,\m)$ and $E^{\eta,3}_{m_1m_2 m_3}(j,J,\m)$ correspond to tensors of mixed symmetry in the decomposition in irreducible representations of $O(N)$.  Tensors with symmetry given by the remaining Young tableau have zero eigenvalue for $\l_t=0$, likewise all the other tensors orthogonal to the ones above, within their same symmetry class.

We find in total $N^3- \f32 N (N+1) +2$ zeros for $N\geq 5$. The $N=3$ case is special because of the different restriction on $J$ in the first set of eigenvectors, and in this case we find 16 zeros.
Surprisingly we have always many more zeros than we would expect from just the symmetry breaking, that is, $\f32 N (N-1)-3$ Goldstone  modes.
The difference is due in large part to the symmetries of the $3jm$ symbol: for $\l_t=0$, two of the $m_i'$ indices of the Hessian belong always to the same $3jm$, hence acting on a tensor with opposed symmetry leads to a null result.

Since the non-zero eigenvalues are  positive, we conclude that for $\l_t=0$ the $SO(3)$-invariant solution is linearly stable, but one should go beyond the linear order in order to asses the fate of the non-Goldstone flat directions.

Turning on $\l_t$ there is a non trivial mixing of the basis vectors, and we have not been able to find the eigenvectors. We have however done some numerical checks for small $N$, and we found that the extra zero modes disappear, leaving us with the expected number of (Goldstone) zero modes.\footnote{Except for $N=7$, where we find 49 zero modes instead of the expected 60. This should be due to an accidental larger symmetry of the solution, which however we could not identify.}
However, some of the new eigenvalues are in general negative, except for specific values of $N$ and limited ranges of $\l_t$. For example, 
for $N=5$ we have found that for $0 \leq \lambda_t \leq \frac{59 \lambda_p}{13 \sqrt{5}}$ the eigenvalues are all non-negative.

\section{Broken phase in the large-$N$ 2PI effective action}
\label{sec:2PI}

 %
We begin with a brief reminder of the 2PI formalism (for further details and references, see Ref.~\cite{Benedetti:2018goh}).
The full 2PI effective action writes
\be
\G [\phi, G] = S [\phi] + \frac{1}{2} \Tr\left[G_0^{-1} * G\right] + \f12 \Tr[\ln G^{-1}] +\G_2 [\phi, G] \,,
\ee
where
\be \label{eq:Gamma_2}
\G_2 [\phi, G] = -\ln \int_{2PI} d\m_{G}[\vph] \; e^{  -  \tilde{S}_{\rm int}[\phi,\varphi] } \,,
\ee
and in the functional integral, $d\m_{G}[\vph]$ is a normalized Gaussian measure with covariance $G_{abc,a'b'c'}$, and the subscript 2PI reminds us that in the perturbative expansion we only retain two-particle irreducible (2PI) diagrams.
The free covariance $G_0$, in the presence of the background field $\phi$,  is defined as
\be \label{eq:brokenG0}
\begin{split}
(G_0^{-1})_{abc,a'b'c'}(x,x') &\equiv \f{\d^2 S}{\d\vph_{abc}(x) \d\vph_{a'b'c'}(x')}[\phi] \\
& = C^{-1}(x,x') \d_{aa'} \d_{bb'}\d_{cc'} +\cH_{abc,a'b'c'}(x) \d(x-x')  \,,
\end{split}
\ee
where $C^{-1}(x,x')$ is the inverse free covariance, i.e.\ integral kernel of the kinetic operator $(-\p_\m\p^\m)^\z$, and the Hessian is defined in Eq.~\eqref{eq:hess1}. 
Traces and $*$ products are in the matrix sense, with a matrix index collectively corresponding to an $O(N)^3$ triplet of indices and a spacetime point.
For example,
\be
\Tr\left[G_0^{-1} * G\right] \equiv \sum_{abc,a'b'c'} \int_{x,y} (G_0^{-1})_{abc,a'b'c'}(x,y) G_{a'b'c',abc}(y,x) \,.
\ee
Lastly, the new interacting action is obtained by expanding the original action around the background $\phi$, and keeping only terms which are cubic or higher in the fluctuation field $\vph$:
\be \label{eq:S-tilde}
\tilde{S}_{\rm int}[\phi,\varphi] \equiv \sum_{n\geq 3} \f{1}{n!} \int_{x_1,\ldots,x_n} \f{\d^n S}{\d\vph_{a_1b_1c_1}(x_1)\cdots \d\vph_{a_n b_n c_n}(x_n)}[\phi] \vph_{a_1b_1c_1}(x_1) \cdots \vph_{a_n b_n c_n}(x_n) \,.
\ee

\begin{figure}
\centering 
\includegraphics[width=0.3\textwidth]{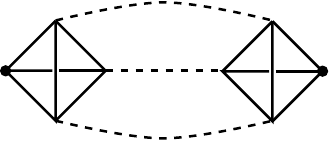}
\hspace{1cm} \includegraphics[width=0.21\textwidth]{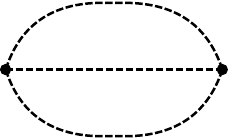}
\caption{\label{fig:melon-bkgr} Example of a two-loop graph (left) with two background fields, and its Feynman diagram (right). In this case we talk of a melonic Feynman diagram.}
\end{figure}

\begin{figure}
\centering 
\includegraphics[width=0.3\textwidth]{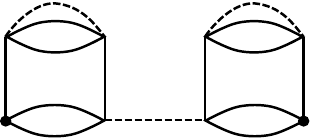}
\hspace{1cm} \includegraphics[width=0.21\textwidth]{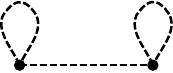}
\caption{\label{fig:pillow-bkgr} Example of a two-loop (one-particle reducible) graph (left) with two background fields, and its Feynman diagram (right). In this case we talk of a melonic graph, and tadpole diagram.}
\end{figure}

In our case, $\tilde{S}_{\rm int}[0,\varphi]$ coincides with the quartic part of  Eq.~\eqref{eq:int-action}, while for $\phi\neq 0$, $\tilde{S}_{\rm int}[\phi,\varphi] $ contains also some cubic terms. The latter have the same graphical representation of the quartic invariants in Eq.~\eqref{eq:int-action-graph}, but with one distinguished vertex, corresponding to the background field, and their couplings are multiplied by a factor 4 with respect to those of the corresponding interactions without background field. We represent the background field with a small disk at a vertex, as before, while we use a vertex without any marker for the fluctuation field. The Wick contraction of two quantum fields is represented by a dotted line, also referred to as ``color 0'' line, and thus after performing all the contractions we end up with 4-colored graphs.
We adopt the following terminology: we use the term ``Feynman diagram'' (or just diagram) in the usual sense, i.e.\ to represent Wick contractions between local interactions in the perturbative expansion of the functional integral (in other words, edges of color 1, 2, and 3 are shrunk to a point, leaving us only with color-0 edges); hence a Feynman diagram only keeps track of the propagators between spacetime points, and the order of the interaction at a given point. For the 4-colored graps, with the contraction pattern of the interactions still explicit, we use instead the term ``combinatorial graph'' of a Feynman diagram, or just graph.
Figures \ref{fig:melon-bkgr} and \ref{fig:pillow-bkgr} show two examples of combinatorial graphs and respective Feynman diagrams, with two interactions having a background field each. The distinction between graphs and diagrams allows us to unambiguously fix the meaning of melonic: melonic diagrams are melonic in spacetime, as used in the SYK literature \cite{Klebanov:2016xxf}; melonic graphs are melonic in their combinatorial structure, as used in earlier tensor model literature \cite{uncoloring}.
Notice that our vacuum Feynman diagrams necessarily contain an even number of background fields, because they always come together with three fluctuation fields, and we need an even number of the latter in a vacuum diagram.

Our goal is to show that in the large-$N$ limit $\G_2[\phi,G]$ is determined by a finite number of diagrams.
In order to achieve that, we need a self-consistent ansatz specifying how $\phi$ and $G$ contribute to the powers of $N$ in a generic diagram appearing in Eq.~\eqref{eq:Gamma_2}. We will base our ansatz on the on-shell behavior, i.e.\ on the solutions of the field equations for  $\phi$ and $G$, which we will partially fix in a self-consistent approach, taking inspiration from the classical solutions.

The one- and two-point functions of the theory are determined by the solution of the field equations
\begin{align} \label{eq:q-eom-phi}
0 & = \f{\d\G}{\d \phi_{abc}} = \f{\d S}{\d \phi_{abc}}  + \frac{1}{2} \Tr\left[\f{\d\cH}{\d \phi_{abc}}  G\right] +\f{\d\G_2}{\d \phi_{abc}} \,, \\
 \label{eq:q-eom-G}
0 & = \f{\d\G}{\d G_{abc,a'b'c'}} = - (G^{-1})_{abc,a'b'c'} +  (G_0^{-1})_{abc,a'b'c'} + 2 \f{\d\G_2}{\d G_{abc,a'b'c'}} \,.
\end{align}
The second equation is nothing but the usual Schwinger-Dyson equation, $G^{-1}= G_0^{-1} - \Sigma$, with the self-energy $\Sigma=-2 \f{\d\G_2}{\d G}$.

The coupled system of equations \eqref{eq:q-eom-phi}-\eqref{eq:q-eom-G} has been considered in the symmetric phase before \cite{Benedetti:2018goh}: the only $O(N)^3$ symmetric solution is $\phi_{abc}=0$ and $G_{abc,a'b'c'} \propto \d_{aa'} \d_{bb'} \d_{cc'}$.
However, it is too complicated to be solved in the broken phase.
We will thus limit ourselves to argue that an $SO(3)$ invariant solution is possible.

First of all, we should transform to discrete Fourier space, as in Sec.~\ref{sec:Fourier}. 
Next, we assume that a solution for the transformed $\phi_{abc}$ and $G_{abc,a'b'c'}$ is invariant under the $SO(3)$ subgroup of \eqref{eq:Fourier-symm} (i.e.\ for rotation matrices \eqref{eq:D-Fourier}, with the same $g\in SO(3)$ for $i=1,2,3$). Then, the on-shell $\phit_{m_1 m_2 m_3}$ must be proportional to a $3jm$ symbol, 
\be \label{eq:phi-assumption}
\phit_{m_1 m_2 m_3} = \hat\Phi_{m_1 m_2 m_3} \equiv  \hat\g\, \tj{j}{j}{j}{m_1}{m_2}{m_3} \,,
\ee
and so must be also the equation of motion, because of the following identity \cite{Yutsis:1962vcy,Varshalovich:1988ye}:
\be \label{eq:3pt-id}
\vcenter{\hbox{\includegraphics[width=1cm]{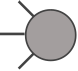}}}\;  = \; \vcenter{\hbox{\includegraphics[width=2.7cm]{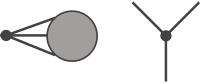}}}\;,
\ee
where the left-hand side represent an arbitrary $SO(3)$ invariant with three free spins (i.e.\ non contracted). Any such invariant is proportional to a $3jm$ symbol, with proportionality constant given by its contraction with the $3jm$ symbol itself.
Hence, the equations of motion reduce to a scalar equation for the proportionality coefficient $\hat\g$:
\be
0 = \sum_{\{m_i\}} \tj{j}{j}{j}{m_1}{m_2}{m_3} \f{\d\G}{\d \phit^{m_1 m_2 m_3}} {\Big|_{\phit=\hat\Phi}} =f(\hat\g;j;\l_2,\l_d,\l_p,\l_t)\,.
\ee
If for some range of couplings $f(\hat\g;j;\l_2,\l_d,\l_p,\l_t)$ has real roots in $\hat\g$, then our assumption is consistent, and an $SO(3)$-invariant solution exists.

The most challenging part is to solve for $G_{abc,a'b'c'}$, as already at the classical level its expression \eqref{eq:brokenG0} is non-trivial, and we do not a priori know the full 2PI effective action.
In order to make progress, we will assume that  the correct scaling behavior in $N$ at leading order is determined by the diagonal part of the two-point function, i.e.\ for the purpose of determining the large-$N$ scaling we will replace
\be \label{eq:G-assumption}
G_{abc,a'b'c'} \to \d_{aa'} \d_{bb'} \d_{cc'} \,.
\ee
Graphically this amounts to the replacement
\be \label{eq:G-assumption-graph}
\vcenter{\hbox{\includegraphics[width=1.9cm]{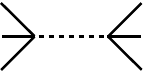}}}\,  \to \; \vcenter{\hbox{\includegraphics[width=1.5cm]{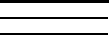}}} \,,
\ee
inside a generic combinatorial graph of a Feynman diagram.
We will use this assumption in order to find the leading-order 2PI effective action.

In a given connected vacuum graph, in the presence of $2n$ background fields $\phi$, the replacement \eqref{eq:G-assumption-graph} leads to a graph with only background vertices, i.e.\ a fully contracted graph of recoupling theory, also known as $3nj$ symbol, with $2n$ vertices and $3n$ lines.\footnote{Our definition of $3nj$ symbol is actually more relaxed than the standard one: generally only 3PI graphs are called $3nj$ symbols, that is, those graphs which are separable in a nontrivial way (i.e.\ into graphs with more than one vertex each) only by cutting more than three lines. This is because otherwise they can be reduced to products of smaller graphs by some standard factorization formulas (namely, Eq.~\eqref{eq:2PR-factoriz} and a similar one for a 3-edge-cut \cite{Yutsis:1962vcy,Varshalovich:1988ye}). We instead call $3nj$ any graph that we obtain as described above, before its reduction by factorization formulas.}
The overall scaling in $N$ of such a graph is a product of three factors: a first factor comes from the explicit scaling of the couplings in the action \eqref{eq:int-action}; a second factor comes from the faces of color $0i$, with $i\in(1,2,3)$, that form after the replacement \eqref{eq:G-assumption-graph}, each face carrying a trace of the identity matrix, i.e.\ a factor $N$; and a third factor comes from the asymptotics of the $3nj$ symbol.
The second factor is relatively well understood: it corresponds to the scaling factor of a $2n$-point function in the symmetric phase, where the external legs are here contracted with the background fields. Schematically, the (zero-dimensional) amplitude of a graph $\cG_{2n,n'}$ with $2n$ interaction bubbles containing a background field, and $n'$ without, is
\be
\begin{split}
A(\cG_{2n,n'}) &\sim N^{-\f32 n_t -2 n_p -3 n_d}\left( \prod_{i=1\ldots 2n} \phi_{a_i b_i c_i} \right) \left\la \left( \prod_{i=1\ldots 2n} \vph_{a_i b_i c_i} \right) \right\ra_{\cG_{2n,n'}}\\
&\sim N^{-\f32 n_t -2 n_p -3 n_d+3n+F(\cG_{2n,n'})}\left( \prod_{i=1\ldots 2n} \phit_{m_1^i m_2^i m_3^i} \right) \prod_{e\in \p(\cG_{2n,n'})} g^{m_e m'_e} \\
&\leq N^{-\f32 n_t -2 n_p -3 n_d+3n+F(\cG_{2n,n'})-\a(\p(\cG_{2n,n'}))}\,,
\end{split}
\ee
where $n_t$, $n_p$, and $n_d$ (with $n_t+n_p+n_d=2n+n'$) are the number of tetrahedron, pillow, and double-trace interactions, respectively, here all taken without background field, which we have instead moved to the external legs. $\la\ldots\ra_{\cG_{2n,n'}}$ is the contribution of $\cG_{2n,n'}$ to the amputated $2n$-point function, $F(\cG_{2n,n'})$ is the number of closed faces in $\cG_{2n,n'}$, and $\p(\cG_{2n,n'})$ is the boundary graph of $\cG_{2n,n'}$, i.e.\ the graph formed by the external fields after Wick contractions, and it corresponds to the $3nj$ symbol. Each edge $e\in \p(\cG_{2n,n'})$ ends on two external fields, with magnetic quantum numbers $m_e,m'_e\in \{m_1^i, m_2^i, m_3^i\}_{i=1\ldots 2n}$, which are contracted by an invariant $SO(3)$ metric, each carrying an extra factor of $N$, due to Eq.~\eqref{eq:delta_m}, for a total factor of $N^{3n}$. Lastly, $\a(\p(\cG_{2n,n'}))$ is the asymptotic scaling factor of the $3nj$ symbol associated to the boundary graph.

For vacuum graphs without background fields, it is well known  \cite{Carrozza:2015adg} that the amplitudes are proportional to $N^{3-\omega}$, with $\omega\geq 0$, and $\omega = 0$ for melon-tadpole diagrams. For combinatorial purposes we can actually forget about pillow and double-trace interactions, as these can be viewed as boundary graphs of graphs with only tetrahedron interactions (e.g.\ \cite{Benedetti:2019eyl}), hence we can consider purely melonic diagrams.
The leading-order graphs for the $2n$ point functions are obtained by cutting open $n$ propagators of leading-order vacuum graphs (corresponding to melonic diagrams), without disconnecting them. In such process we loose a number $F_L(\cG_{2n,n'})$ of faces, and hence we loose $F_L(\cG_{2n,n'})$ factors of $N$. Therefore, we have 
\be
-\f32 n_t -2 n_p -3 n_d+F(\cG_{2n,n'}) = 3-\omega -F_L(\cG_{2n,n'})\,.
\ee

Furthermore, for $2n$-point functions obtained from melonic vacuum graphs we have
\be
F_L(\cG^{\rm melonic}_{2n,n'})=2n+c(\p(\cG^{\rm melonic}_{2n,n'}))\,,
\ee
where $c(\p(\cG_{2n,n'}))$ is the number of connected components of the boundary graph of $\cG_{2n,n'}$. Therefore, we have
\be
A(\cG^{\rm melonic}_{2n,n'})  \sim N^{3-(2n+c(\p(\cG^{\rm melonic}_{2n,n'})))+3n-\a(\p(\cG^{\rm melonic}_{2n,n'}))} \,.
\ee
At this stage we use the fact that a connected melonic $3nj$ symbol is equal to $N^{-n+1}$ \cite{Amit:1979ev}. 
This can be easily proven by repeated use of the following identity:
\be \label{eq:2PR-factoriz}
\vcenter{\hbox{\includegraphics[width=2.2cm]{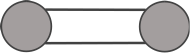}}}\;  = \f1N \; \vcenter{\hbox{\includegraphics[width=2.2cm]{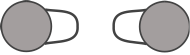}}}\;.
\ee
For a $\cG_{2n,n'}$ obtained from a melonic vacuum graph, we thus have $\a(\p(\cG^{\rm melonic}_{2n,n'})) = n-c(\p(\cG^{\rm melonic}_{2n,n'}))$, and we obtain
\be
A(\cG^{\rm melonic}_{2n,n'})  \sim N^{3} \,.
\ee

To the best of our knowledge, the general asymptotic behavior of $3nj$ symbols is an open problem,\footnote{See for example Ref.~\cite{Costantino2011,Bonzom:2011cy,Dona:2017dvf} and references therein.} but in Ref.~\cite{Amit:1979ev} it was also argued (by a mix of analytical and numerical arguments) that non-melonic $3nj$ symbols are subleading: $|(3nj)_{\rm non-melonic}|\leq N^{-n+1-\a'}$ for $N=(2j+1)\to\infty$, with $\a'>0$.

We conclude that leading-order graphs are obtained by cutting propagators of graphs corresponding to melonic diagrams, without disconnecting them, and placing background fields on the endpoints of removed propagators. The resulting boundary graph defines a melonic $3nj$ symbol.
Combining this with the 2PI condition on the diagrams contributing to Eq.~\eqref{eq:Gamma_2}, we infer that, besides the graphs without background fields already discussed in Ref.~\cite{Benedetti:2018goh,Benedetti:2019eyl}, we have only one new graph, the one in Fig.~\ref{fig:melon-bkgr}.

At last, under the hypotheses that we have made above, the resulting 2PI effective action is:
\be \label{eq:2PI-result}
\begin{split}
\G_2 [\phi, G] = & \f{\l_d}{4N^3} \int_x \left(G_{a_1 a_2 a_3,a_1 a_2 a_3}(x,x)\right)^2 \\
&+\f{\l_p}{4N^3} \sum_i \a_i \d_{a_ia'_i}\d_{b_ib'_i} (\prod_{j\neq i} \d_{a_jb_j} \d_{a'_jb'_j})\int_x G_{a_1 a_2 a_3,b_1 b_2 b_3}(x,x) G_{a'_1 a'_2 a'_3, b'_1 b'_2 b'_3}(x,x)\\
&-\f{\l_t^2}{8 N^3} \int_{x,y}G_{a_1 a_2 a_3, b_1 b_2 b_3}(x,y) G_{a_1 a'_2 a'_3, b_1 b'_2 b'_3}(x,y)  G_{a'_1 a_2 a'_3, b'_1 b_2 b'_3}(x,y) G_{a'_1 a'_2 a_3, b'_1 b'_2 b_3}(x,y)\\
&-\f{\l_t^2}{2 N^3} \int_{x,y} \phi_{a_1 a_2 a_3}(x)\phi_{b_1 b_2 b_3}(y) G_{a_1 a'_2 a'_3, b_1 b'_2 b'_3}(x,y)  G_{a'_1 a_2 a'_3, b'_1 b_2 b'_3}(x,y) G_{a'_1 a'_2 a_3, b'_1 b'_2 b_3}(x,y) \,.
\end{split}
\ee
Ideally, we should check the consistency of our hypotheses by solving the field equations, and showing that indeed the leading-order scaling in $N$ is captured by the diagonal part of $G$. Unfortunately, solving the equations of motion is very hard.
What we can straightforwardly do, is to check that with the assumptions \eqref{eq:phi-assumption} and \eqref{eq:G-assumption} all the terms in the 2PI effective action, including in particular the last term in Eq.~\eqref{eq:2PI-result}, are indeed of order $N^3$.

\section{Conclusions}

In this paper we have identified an $SO(3)$-invariant solution to the field equations of the bosonic $O(N)^3$-invariant tensor model with quartic interactions. We have highlighted the similarity between the graphical representation of tensor invariants and that of $SU(2)$ recoupling theory, after a discrete Fourier transform on the tensor indices, making the appearance of such solution natural even at a quantum or statistical level, that is, in the full quantum effective action. For the latter, we have employed the 2PI version, which is well suited to the large-$N$ limit of vector and tensor models \cite{Benedetti:2018goh}, but which is of course equivalent to the more usual 1PI effective action, once the Schwinger-Dyson equations for the two-point function are used.

An interesting feature of the $SO(3)$-invariant solution is that it satisfies the scaling relations in equations (\ref{eq:phi-eq-1}-\ref{eq:phi-eq-3}), implying that all three quartic interactions contribute at the same order in $N$ to the action and the field equations. In order to fully appreciate this point, we study in App.~\ref{app:othersol} other possible solutions, which however either do not satisfy such scalings, and are unstable, or are possible only for a reduced set of interactions.
The scalings in Eq.~(\ref{eq:phi-eq-1}-\ref{eq:phi-eq-3}) play an essential role in determining the large-$N$ form of the 2PI effective action in Eq.~\eqref{eq:2PI-result}, therefore we believe that the $SO(3)$-invariant solution plays an important role in these models.

In this respect, we make the following observation on the relation between the Amit-Roginsky model and the SYK model. The Amit-Roginsky model, introduced in Ref.~\cite{Amit:1979ev}, is a bosonic model of fields in an irreducible representation of $SO(3)$, and with a cubic interaction mediated by a $3jm$ symbol. The action reads:
\be \label{eq:AR}
\begin{split}
S_{\rm AR}= \int d^d x \; &\left(\sum_m (\p_\n\bar\phi_m \p^\n\phi^m + \m^2 \bar\phi_m \phi^m) \right.\\
&\left.+\sum_{m_1,m_2,m_3} \f{g}{3!}\sqrt{2j+1}\tj{j}{j}{j}{m_1}{m_2}{m_3}( \phi^{m_1}\phi^{m_2}\phi^{m_3} + \bar\phi_{m_1}\bar\phi_{m_2}\bar\phi_{m_3} ) \right) \,,
\end{split}
\ee
where indices are raised by the invariant metric, as in Sec.~\ref{sec:Fourier}.
The model is solvable in the large-$j$ limit, as it leads to melonic dominance (on  the basis of the conjecture we discussed in Sec.~\ref{sec:2PI}).
We then observe that in the light of our results, we could view the Amit-Roginsky model as an on-shell version (or saddle-point approximation) of a bosonic SYK-style model with quenched disorder \cite{Sachdev:1992fk,Kitaev,Polchinski:2016xgd,Maldacena:2016hyu}, i.e.\ with a randomly distributed rank-3 tensor coupling in place of the $3jm$ symbol. The distribution would need to be non-Gaussian, and with at least a pillow or tetrahedron quartic term and a negative coupling for the quadratic term, in order to allow a non-trivial solution, but that does not lead to crucial differences with respect to the Gaussian case, as shown in Ref.~\cite{Krajewski:2018lom}.

Several possible generalizations of our results are conceivable. A first obvious one is the extension of our solution to higher-order potentials, as for example the sextic potentials considered in Ref.~\cite{Giombi:2018qgp,Benedetti:2019rja}. Based on the discussion in Sec.~\ref{sec:2PI}, it seems evident that a similar solution could be possible for much more general potentials.
A second generalization would be to investigate the possibility of having $SO(n)$-invariant solutions with $3<n<N$.\footnote{Notice that no $O(N)$-invariant tensor exists for $N>3$ \cite{jeffreys_1973} (for $N=3$ we can of course have $\phi_{abc} \propto \eps_{abc}$, the Levi-Civita symbol, which is invariant under $SO(3)$, and which is in fact related to our solution at $N=3$).}
In fact, $3mj$ symbols can be constructed also for larger groups \cite{Hamermesh}, and they can probably be used to construct solutions at least for $\l_t=0$, as in such case we only need the orthogonality relations in Eq.~\eqref{eq:phi-eq-1}, which are generically satisfied by $3mj$ symbols.
An explicit example could be constructed with $SO(4)$, for which one can use the fact that it is isomorphic to $SU(2)\times SU(2)$ \cite{Wybourne}. However, for non-simply-reducible groups additional labels are required, and at the moment it is not clear to us how that would affect our construction.

\section*{Acknowledgements}

We would like to thank Valentin Bonzom, Simone Speziale, and Massimo Taronna for useful  discussions.

IC acknowledges support through the Erasmus+ programme and hospitality at the Laboratoire de Physique Th\'eorique d'Orsay (Univ.Paris-Sud) during the first stages of this work.

The work of DB is supported by the European Research Council (ERC) under the European Union's Horizon 2020 research and innovation program (grant agreement No818066).

\newpage

 \appendix

\section{$3jm$ and $6j$ symbols: some formulas and conventions}
\label{app:3j}

We mostly follow the notation and conventions of Ref.~\cite{Yutsis:1962vcy} (see also Ref.~\cite{Varshalovich:1988ye,Brink:1993}).

\paragraph{$3jm$ symbol.}
Wigner's $3jm$ symbol is defined in terms of the Clebsch-Gordan coefficients:
\be
\tj{j_1}{j_2}{j_3}{m_1}{m_2}{m_3} = \f{(-1)^{j_1-j_2+j_3}}{\sqrt{2j_3+1}} (-1)^{j_3-m_3} \la j_3 -m_3 | j_1 m_1, j_2 m_3\ra \,.
\ee
The $3jm$ symbol is non-zero only if the angular momenta (or spin) $j_1$, $j_2$, and $j_3$ satisfy the triangular inequality $|j_1-j_2|\leq j_3 \leq j_1+j_2$ and their sum is an integer, and if their projections (or magnetic quantum numbers) sum up to zero:
\be \label{eq:m0}
m_1+m_2+m_3=0 \,.
\ee
In this paper we are mostly interested in the case $j_1=j_2=j_3=j$ (and thus integer $j$), for which the conditions on the angular momenta are obviously satisfied, but we also use $3jm$ symbols with different spins for the analysis of the Hessian in Sec.~\ref{sec:Fourier}.
Sometimes the $3jm$ symbols are called simply $3j$, but we reserve this name for the triangular inequality symbol, that is, we define a $3j$ symbol $ \{j_1 j_2 j_3\}$ as being equal to one if $j_1$, $j_2$, and $j_3$ satisfy the triangular inequality, and zero otherwise.

The $3jm$ symbols have the following symmetries:
\begin{enumerate}
\item $SU(2)$ invariance:
\be \label{eq:3j-invariance}
\tj{j_1}{j_2}{j_3}{m_1}{m_2}{m_3}  = \sum_{m'_1,m'_2,m'_3}\tj{j_1}{j_2}{j_3}{m'_1}{m'_2}{m'_3}  D^{j_1}_{m_1 m'_1}(g) D^{j_2}_{m_2m'_2}(g) D^{j_3}_{m_3m'_3}(g) \,,
\ee
where $D^{j}_{mm'}(g)$ is a Wigner's $D$-matrix, i.e.\ the matrix corresponding to $g\in SU(2)$ in the representation labelled by $j$.
For integer $j$, these are representations of $SO(3)$, hence the $3jm$ of our interest is an $SO(3)$ invariant.
\item Permutation symmetries:
\be \label{eq:perm1}
\tj{j_1}{j_2}{j_3}{m_1}{m_2}{m_3} = \tj{j_2}{j_3}{j_1}{m_2}{m_3}{m_1} = \tj{j_3}{j_1}{j_2}{m_3}{m_1}{m_2} \,,
\ee
\be \label{eq:perm2}
\tj{j_1}{j_2}{j_3}{m_1}{m_2}{m_3} = (-1)^{j_1+j_2+j_3}\tj{j_2}{j_1}{j_3}{m_2}{m_1}{m_3}
  = (-1)^{j_1+j_2+j_3}\tj{j_1}{j_3}{j_2}{m_1}{m_3}{m_2}  \,.
\ee

\item Time-reversal:
\be \label{eq:time-rev}
\tj{j_1}{j_2}{j_3}{m_1}{m_2}{m_3} = (-1)^{j_1+j_2+j_3} \tj{j_1}{j_2}{j_3}{-m_1}{-m_2}{-m_3}  \,.
\ee
\item Regge symmetries, which we do not need here.
\end{enumerate}
Notice that in the case $j_1=j_2=j_3=j$, the $3jm$ symbol is fully symmetric (antisymmetric) under permutations and time-reversal if $j$ is even (odd).

We introduce also the invariant metric $g_{j}^{m m'} $, defined as:
\be \label{eq:metric_j}
g_{j}^{m m'} = g^{j}_{m m'} \equiv \sqrt{2j+1} \tj{j}{j}{0}{m}{m'}{0} = (-1)^{j-m} \d_{m\, -m'} \,,
\ee
which is also its own inverse, i.e.\ $\sum_{m''} g^{j}_{m m''} g_{j}^{m'' m'} = \d_m^{m'}$. 

A graphical calculus is introduced by representing the $3jm$ symbol as a three-valent vertex. Edges joining pairs of vertices denote a contraction, i.e.\ a summation over pairs of magnetic quantum numbers $m_i$ associated to equal spins $j_i$, with a weight $(-1)^{j_i-m_i}$ and with the two magnetic quantum numbers appearing with opposite signs. In other words, the contraction is performed with the invariant metric \eqref{eq:metric_j}, which we represent by a line without vertices whenever it appears explicitly.
By taking products of $3jm$ symbols and contracting them in different ways, we obtain new objects that depend on the free spins and their projections, as well as on the contracted spins (whose projections are summed over). Such objects are called $jm$ coefficients. In the special case in which the number of contracted spins, for a given number of free spins, is minimized, they are also called generalized Wigner coefficients. From a graphical point of view, generalized Wigner coefficients correspond to tree diagrams.

Among other formulas of importance to us is the orthogonality relation:
\be \label{eq:orthog1}
\sum_{m_2, m_3} \tj{j_1}{j_2}{j_3}{m_1}{m_2}{m_3} \tj{j'_1}{j_2}{j_3}{m'_1}{m_2}{m_3} = \f{1}{2j_1+1} \d_{j_1 j'_1}\d_{m_1 m'_1} \{j_1 j_2 j_3\} \,,
\ee
which, using Eq.~\eqref{eq:time-rev} and \eqref{eq:m0}, can also be written as
\be
\begin{split}
\sum_{m_2, m_3} & (-1)^{j_2-m_2+j_3-m_3} \tj{j_1}{j_2}{j_3}{m_1}{m_2}{m_3} \tj{j'_1}{j_2}{j_3}{-m'_1}{-m_2}{-m_3} \\
&= \f{(-1)^{j_1-m_1}}{2j_1+1} \d_{j_1 j'_1}\d_{m_1 m'_1} \{j_1 j_2 j_3\} \,,
\end{split}
\ee
and represented as\footnote{In order to not overburden the drawings, we generally omit (some of) the labels in the graphical representations, as well as the orientations of the vertices, if they are obvious or deducible from the explicit formula which we provide with them. In fact, as in the main body of the paper, our graphs are only meant for a better visualization of the formulas.}
\be\nn
j_1\vcenter{\hbox{\includegraphics[width=1.8cm]{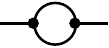}}}j'_1 = \f{\d_{j_1 j'_1} }{2j_1+1} \vcenter{\hbox{\includegraphics[width=1.5cm]{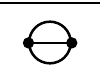}}} \, ,
\ee
where on the right-hand side we have introduced also the melon graph corresponding to the identity
\be
\sum_{m_1,m_2, m_3} (-1)^{\sum_{i=1\ldots 3}(j_i-m_i)}\tj{j_1}{j_2}{j_3}{m_1}{m_2}{m_3} \tj{j_1}{j_2}{j_3}{-m'_1}{-m_2}{-m_3} =  \{j_1 j_2 j_3\} \,.
\ee

If instead we sum over only one magnetic number, but also on the respective spin, we have the equality
\be
\sum_{J, m} (-1)^{J-m}(2J+1) \tj{j_1}{j_2}{J}{m_1}{m_2}{m} \tj{j_1}{j_2}{J}{m'_1}{m'_2}{-m} 
= g_{j_1}^{m_1 m'_1} g_{j_2}^{m_2 m'_2}   \,,
\ee
which is represented as
\be
\sum_{J} (2J+1)  \vcenter{\hbox{\includegraphics[width=1.8cm]{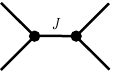}}} = \, \vcenter{\hbox{\includegraphics[width=1.5cm]{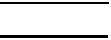}}} \, .
\ee

Notice that the relation stating the invariance of the metric,
\be \label{eq:g-invar}
\sum_{m_1,m_2}g_j^{m_1 m_2} D^{j}_{m m_1}(g) D^{j}_{m' m_2}(g)  = g_j^{m m'} \,,
\ee
is as close as we get to orthogonality of the $D$-matrices, which, for integer $j$, form a unitary representation of $SO(3)$:
\be \label{eq:D-unitarity}
\sum_{m_1}  D^{j}_{m m_1}(g) D^{j}_{m' m_1}(g)^*  = \d_{m m'} \,.
\ee
Equations \eqref{eq:g-invar} and \eqref{eq:D-unitarity} are related by the the identity
\be \label{eq:Dconj}
D^{j}_{m m'}(g)^* = (-1)^{2j-m-m'} D^{j}_{-m -m'}(g) \,.
\ee

In view of the definition \eqref{eq:metric_j}, we have as a special case of Eq.~\eqref{eq:orthog1} the following tracelessness condition for the $3jm$ symbol with equal angular momenta (for $j\neq 0$):
\be
\sum_{m_2, m_3} \tj{j}{j}{j}{m_1}{m_2}{m_3} g_j^{m_2 m_3} = 0\,.
\ee

\paragraph{$6j$ symbol.}
The $6j$ symbol can be defined in terms of $3jm$ symbols via the relation
\be \label{eq:3j-triangle}
\begin{split}
\sj{j_1}{j_2}{j_3}{j'_1}{j'_2}{j'_3} & = \sum_{\{m\},\{m'\}} (-1)^{\sum_{i=1\ldots 3} (j'_i- m'_i+j_i-m_i)}  \tj{j'_1}{j'_2}{j_3}{m'_1}{-m'_2}{m_3} \\
&\qquad\quad \times \tj{j'_2}{j'_3}{j_1}{m'_2}{-m'_3}{m_1} \tj{j'_3}{j'_1}{j_2}{m'_3}{-m'_1}{m_2} \tj{j_1}{j_2}{j_3}{m_1}{m_2}{m_3}\\
& = \, \vcenter{\hbox{\includegraphics[width=1.5cm]{tetrahedron-2.pdf}}} \, ,
\end{split}
\ee
and it is invariant under permutations of its columns or under interchange of the upper and lower spins in each of any two columns.
We have omitted the spin labels in the graphical representation, but it can be easily reconstructed by the following rule: the spins in the first row of the $6j$ symbol share the same vertex, and spins in the same column do not share any vertex.

There are three formulas relating products of $3jm$ symbols to $6j$ symbols that are particularly useful for us.
The first involves a product of two $3jm$ symbols:
\be \label{eq:t-channel}
\begin{split}
\sum_{m_3} & (-1)^{j_3-m_3} \tj{j_1}{j_2}{j_3}{m_1}{m_2}{m_3} \tj{j'_1}{j'_2}{j_3}{m'_1}{m'_2}{-m_3}\\
 & = \sum_{J, m} (-1)^{J-m}(2J+1) \tj{j_1}{j'_2}{J}{m_1}{m'_2}{m} \tj{j'_1}{j_2}{J}{m'_1}{m_2}{-m}  \sj{j_2}{j'_1}{J}{j'_2}{j_1}{j_3}  \,,
\end{split}
\ee
or
\be
 \vcenter{\hbox{\includegraphics[width=2.2cm]{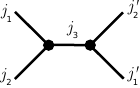}}} = \sum_{J} (2J+1) \; \vcenter{\hbox{\includegraphics[width=1.4cm]{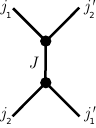}}}\; \vcenter{\hbox{\includegraphics[width=1.5cm]{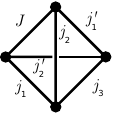}}} \, .
\ee
The second applies to a product of three $3jm$ symbols:
\be \label{eq:3j-triangle}
\begin{split}
\sum_{m'_1, m'_2, m'_3} &(-1)^{\sum_{i=1\ldots 3} (j'_i- m'_i)}  \tj{j'_1}{j'_2}{j_3}{m'_1}{-m'_2}{m_3} \tj{j'_2}{j'_3}{j_1}{m'_2}{-m'_3}{m_1} \tj{j'_3}{j'_1}{j_2}{m'_3}{-m'_1}{m_2} \\
&= \tj{j_1}{j_2}{j_3}{m_1}{m_2}{m_3} \sj{j_1}{j_2}{j_3}{j'_1}{j'_2}{j'_3} \, ,
\end{split}
\ee
or
\be
\vcenter{\hbox{\includegraphics[angle=-90,width=2.1cm]{tetrahedron-eom.pdf}}} \;  = \; \vcenter{\hbox{\includegraphics[width=1.cm]{mass-eom.pdf}}} 
\;  \vcenter{\hbox{\includegraphics[width=1.5cm]{tetrahedron-2.pdf}}} \;,
\ee
which is a special case of Eq.~\eqref{eq:3pt-id}.

And the last one to a product of four $3jm$ symbols:
\be \label{eq:3j-square}
\begin{split}
\sum_{m'_1, m'_2, m'_3} &(-1)^{\sum_{i=1\ldots4} (j'_i- m'_i)}  \tj{j'_1}{j_1}{j'_2}{m'_1}{m_1}{-m'_2}  \tj{j'_2}{j_2}{j'_3}{m'_2}{m_2}{-m'_3}\\
&\qquad\quad \times \tj{j'_3}{j_3}{j'_4}{m'_3}{m_3}{-m'_4}  \tj{j'_4}{j_4}{j'_1}{m'_4}{m_4}{-m'_1} \\
&= (-1)^{j'_4-j_1-j_4-j'_2} \sum_{J, m} (-1)^{J-m}(2J+1) \tj{j_1}{J}{j_4}{m_1}{m}{m_4} \tj{j_2}{J}{j_3}{m_2}{-m}{m_3} \\
&\qquad\quad \times \sj{j_1}{J}{j_4}{j'_4}{j'_1}{j'_2} \sj{j_2}{J}{j_3}{j'_4}{j'_3}{j'_2} \,,
\end{split}
\ee
or
\be
 \vcenter{\hbox{\includegraphics[width=2cm]{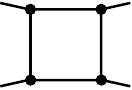}}} = \sum_{J} (2J+1)  \vcenter{\hbox{\includegraphics[width=1.8cm]{2-3jm-t-channel.pdf}}} \;  \vcenter{\hbox{\includegraphics[width=1.5cm]{tetrahedron-2.pdf}}}\;  \vcenter{\hbox{\includegraphics[width=1.5cm]{tetrahedron-2.pdf}}} \;.
\ee

Lastly, another useful formula for us is \cite{Varshalovich:1988ye}:
\be \label{eq:two6j}
\sum_{K} (2K+1) \sj{j_1}{j_2}{K}{j_3}{j_4}{L} \sj{j_1}{j_2}{K}{j_3}{j_4}{L'} = \f{\d_{LL'}}{(2L+1)} \{j_1 j_4 L\} \{j_2 j_3 L\}\,.
\ee
%

\section{Other solutions and their stability}
\label{app:othersol}

We will now consider a few different possible patterns of symmetry breaking and we will verify if each of the solutions is a minimum for the action or not by studying the eigenvalues of the Hessian. The latter is given by:
\be \label{eq:hess1}
\begin{split}
	\cH_{abc,a'b'c'}(x)\equiv & \int_{x'}\f{\d^2S_{\rm int}}{\d \phi_{abc}(x)\d\phi_{a'b'c'}(x')} \\
	=&\,\l_2 \d_{aa'}\d_{bb'}\d_{cc'}+\f{\l_d}{N^3}(2\phi_{abc}\phi_{a'b'c'}+\phi_{a''b''c''}\phi_{a''b''c''}\d_{aa'}\d_{bb'}\d_{cc'})\\&
	+\f{\l_p}{N^2}[\a_1(\d_{aa'}\phi_{a''bc}\phi_{a''b'c'}+\phi_{a'bc}\phi_{ab'c'}+\phi_{a'b''c''}\phi_{ab''c''}\d_{bb'}\d_{cc'})\\&
	+\a_2(\phi_{ab''c}\phi_{a'b''c'}\d_{bb'}+\phi_{ab'c}\phi_{a'bc'}+\phi_{a''b'c''}\phi_{a''bc''}\d_{aa'}\d_{bb'})\\&
	+\a_3(\phi_{abc''}\phi_{a'b'c''}\d_{cc'}+\phi_{abc'}\phi_{a'b'c}+\phi_{a''b''c}\phi_{a''b''c'}\d_{aa'}\d_{bb'})]\\&
	+\f{\l_t}{N^{3/2}}(\phi_{a''b'c}\phi_{a''bc'}\d_{aa'}+\phi_{a'b''c}\phi_{ab''c'}\d_{bb'}+\phi_{a'bc''}\phi_{ab'c''}\d_{cc'})\,,
\end{split}
\ee
where we have integrated over $x'$ in order to get rid of an overall delta function. In the following we will omit the $x$-dependence of $\cH$, as we are interested in the constant solutions.
The Hessian has the following graphical representation:
\be
\begin{split}
\cH_{abc,a'b'c'} = & \, \l_2 \,\vcenter{\hbox{\includegraphics[width=1.2cm]{Hess-mass.pdf}}}
+ \f{\l_d}{N^3} \, \left(2\, \vcenter{\hbox{\includegraphics[width=1.2cm]{Hess-disconnect-1.pdf}}} +\vcenter{\hbox{\includegraphics[width=2cm]{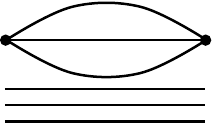}}}  \right) \\
& + \f{\l_p}{N^2}  \sum_{i=1,2,3} \a_i \,\left(\vcenter{\hbox{\includegraphics[width=1.2cm]{Hess-pillow-1.pdf}}}  + \vcenter{\hbox{\includegraphics[width=1.4cm]{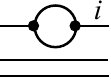}}} + \vcenter{\hbox{\includegraphics[width=.8cm]{Hess-pillow-3.pdf}}}  \right) \\
& +\f{\l_t}{N^{3/2}} \sum_{i=1,2,3} \, \vcenter{\hbox{\includegraphics[width=1.2cm]{Hess-tetra.pdf}}} \,,
\end{split}
\ee
where open half-edges on the left (right) correspond to the unprimed (primed) free indices, and the index label is indicated only where necessary.

\subsection{$O(N^3-1)$-invariant solutions for  $\l_p=\l_t=0$}

For $\l_p=\l_t=0$, the internal symmetry of the model is enhanced to $O(N^3)$. In this case, the model is indeed a vector model in disguise, as it is easily recognized by a collective index notation such as $abc\to{\bf a}=1,\ldots,N^3$.
It follows that for $\l_2<0$ we find the usual symmetry breaking solutions as in an $O(N^3)$-invariant vector model, with stabilizer given by the subgroup of rotations in the hyperplane orthogonal to the non-zero vector solution.
Notice that such a solution will necessarily be such that $\phi_{abc}\phi_{abc} \sim N^3$, which is rather natural for a vector with $N^3$ components of order one, and it implies that the action evaluated on-shell scales like $N^{3}$. 
Furthermore, it is easily checked that for such solution the on-shell action is negative, and hence such stationary point is the absolute minimum (the action is bounded from below, it grows at infinity, and the $\phi_{abc}=0$ stationary point has zero action).

\subsection{$O(N-1)^{3}$-invariant solutions}

Let us suppose that the solution of Eq.~\eqref{eq:eom} can be written as:
\be \label{eq:vector-sol}
\phi_{abc} = u_a v_b w_c \,,
\ee
for three independent vectors $u$, $v$, and $w$ (color symmetry would impose $u=v=w$).
In such case, $O(N-1)^{3}$ is clearly the stabilizer.
Upon substitution of Eq.~\eqref{eq:vector-sol} in Eq.~\eqref{eq:eom} we obtain the equation
\be\label{eq:eom111}
0 = \l_2 + (\f{\l_d}{ N^3} + \f{\l_p (\a_1+\a_2+\a_3) }{N^2} +  \f{\l_t}{N^{3/2}} ) u^2 v^2 w^2 \,,
\ee
which admits real solutions for positive couplings if $\l_2<0$, as anticipated:
\be
u^2v^2w^2=-\f{\l_2N^3}{\l_d+\l_pN(\a_1+\a_2+\a_3)+N^{3/2}\l_t}.
\ee
 Solutions are also possible for $\l_2>0$, if the combination of couplings appearing in the denominator is negative.

Such type of solution is problematic in the large-$N$ limit, because the three types of interactions scale identically with $N$, but due to the different rescaling of their couplings which we have chosen in Eq.~\eqref{eq:action}, we end with a non-homogeneous equation,  as in Eq.~\eqref{eq:eom111}.
In order to take the large-$N$ limit in the field equations we could scale the vectors such that $u^2 v^2 w^2 \sim N^{3/2}$, so that the limit is finite (and the solution non-trivial), but then only $\l_t$ will contribute; alternatively, we could rescale the couplings so that each interaction comes with the same power of $N$ in front.
On the other hand, the scaling of the couplings in Eq.~\eqref{eq:action} was chose because it is the unique one that leads to all three interactions contributing at the same leading order in the Feynman diagrams of the model, and we wish to understand classical solutions that might play a role in that context.

Notice that the action evaluated on-shell (with the scaling that allows a non-trivial large-$N$ limit of the field equations) scales like $N^{3/2}$. Furthermore, for $\l_2<0$ and $\l_t>0$, the action is negative, as in the vector case. On the other hand, for $\l_2>0$ and  $\l_t<0$, the action is positive, hence such non-trivial solution is subdominant with respect to the trivial one.\\
If we substitute Eq.~\eqref{eq:vector-sol} in Eq.~\eqref{eq:hess1}, we obtain:
\be
\begin{split}
\cH_{abca'b'c'}&=\left(\l_2-\f{\l_2\l_d}{\l_d+\l_pN(\a_1+\a_2+\a_3)+N^{3/2}\l_t}\right)\d_{aa'}\d_{bb'}\d_{cc'}\\&+\left[\f{2\l_d}{N^3}
+\f{\l_p}{N^2}(\a_1+\a_2+\a_3)\right] u_au_{a'}v_bv_{b'}w_cw_{c'}\\&+\left(\f{\a_1\,\l_p}{N^2}+\f{\l_t}{N^{3/2}}\right)u^2\d_{aa'}v_bv_{b'}w_cw_{c'}\\&
+\left(\f{\a_2\,\l_p}{N^2}+\f{\l_t}{N^{3/2}}\right)v^2u_au_{a'}\d_{bb'}w_cw_{c'}\\&
+\left(\f{\a_2\,\l_p}{N^2}+\f{\l_t}{N^{3/2}}\right)w^2u_au_{a'}v_bv_{b'}\d_{cc'}\\&
+\f{\l_p}{N^2}(\a_1v^2w^2u_au_{a'}\d_{bb'}\d_{cc'}+\a_2u^2w^2\d_{aa'}v_bv_{b'}\d_{cc'}+\a_3u^2v^2\d_{aa'}\d_{bb'}w_cw_{c'}).
\label{eq:hess2}
\end{split}
\ee
To find the eigenvectors, and then the eigenvalues of this Hessian we can first of all notice that Eq.~\eqref{eq:hess2} has the following form:
\be
\cH_{abca'b'c'}=\sum_{i}M^{1i}_{aa'}M^{2i}_{bb'}M^{3i}_{cc'},\,\,\, \text{with}\,\, M^{1i}_{aa'}=A^{1i}\d_{aa'}+B^{1i}u_au_{a'}.
\ee
This implies that the eigenvectors of $\cH$ are given by products of the eigenvectors of each matrix $M$. The latter are given by the set composed by the vectors with which we have decomposed $\phi_{abc}$ and the ones orthogonal to them. For example, the eigenvectors of each matrix $M^{1i}$ are given by the set $(u,u^{\bot})$ and since $u$ identifies a direction in vector space $\mathbb{R}^N$, we can find $N-1$ eigenvectors of the type $u^{\bot}$.\\
The possible eigenvectors of the Hessian are then of the form:
\be
uvw\,\,\,uv^{\bot}w\,\,\,uv^{\bot}w^{\bot}\,\,\,uvw^{\bot}\,\,\,u^{\bot}v^{\bot}w^{\bot}\,\,\,u^{\bot}v^{\bot}w\,\,\,u^{\bot}vw^{\bot}\,\,\,u^{\bot}vw.
\ee
These are a total of $N^3$  orthogonal tensors, as they should in order to form a basis.
One can then compute the eigenvalues easily and find:
\be
\begin{split}
&\cH_{abca'b'c'}(u_{a'}v_{b'}w_{c'})=-\l_2\left(\f{2\l_d+\l_pN(\a_1+\a_2+\a_3)+2\l_tN^{3/2}}{\l_d+\l_pN(\a_1+\a_2+\a_3)+\l_tN^{3/2}}\right)u_av_bw_c \,,\\&
\cH_{abca'b'c'}(u_{a'}v^{\bot}_{b'}w_{c'})=0 \,,\\&
\cH_{abca'b'c'}(u_{a'}^{\bot}v_{b'}w_{c'})=0 \,,\\&
\cH_{abca'b'c'}(u_{a'}v_{b'}w_{c'}^{\bot})=0 \,,\\&
\cH_{abca'b'c'}(u_{a'}v^{\bot}_{b'}w^{\bot}_{c'})=\l_2\left(\f{\l_pN(\a_2+\a_3)+\l_tN^{3/2}}{\l_d+\l_pN(\a_1+\a_2+\a_3)+\l_tN^{3/2}}\right)u_av^{\bot}_bw^{\bot}_c \,,\\&
\cH_{abca'b'c'}(u_{a'}^{\bot}v_{b'}w^{\bot}_{c'})=\l_2\left(\f{\l_pN(\a_1+\a_3)+\l_tN^{3/2}}{\l_d+\l_pN(\a_1+\a_2+\a_3)+\l_tN^{3/2}}\right)u^{\bot}_av_bw^{\bot}_c \,,\\&
\cH_{abca'b'c'}(u_{a'}^{\bot}v^{\bot}_{b'}w_{c'})=\l_2\left(\f{\l_pN(\a_1+\a_2)+\l_tN^{3/2}}{\l_d+\l_pN(\a_1+\a_2+\a_3)+\l_tN^{3/2}}\right)u^{\bot}_av^{\bot}_bw_c \,,\\&
\cH_{abca'b'c'}(u^{\bot}_{a'}v^{\bot}_{b'}w^{\bot}_{c'})=\l_2\left(\f{\l_pN(\a_1+\a_2+\a_3)+\l_tN^{3/2}}{\l_d+\l_pN(\a_1+\a_2+\a_3)+\l_tN^{3/2}}\right)u^{\bot}_av^{\bot}_bw^{\bot}_c \,.
\end{split}
\ee
	We see that for $\lambda_t>0$ and $\lambda_2<0$ the obtained eigenvalues have mixed signs and that eigenvectors of the type $uv^{\bot}w$ have null eigenvalues.  The presence of negative eigenvalues indicates that this stationary point is unstable. We can show that the null mass eigenvalues are Goldstone's modes by making an infinitesimal rotation of the tensor $\varphi_{abc}$: 
\be
\phi_{abc}\rightarrow\tilde{\phi}_{abc}\simeq\phi_{abc}+\th_{\a}L^{\a}_{aa'}\phi_{a'bc}+\tau_{\a}L^{\a}_{bb'}\phi_{ab'c}+\omega_{\a}L^{\a}_{cc'}\phi_{abc'},\,\,\, \text{with}\, \a=1,\dots,\f{N(N-1)}{2}.
\label{phivar}
\ee
From Goldstone's theorem we know that symmetry transformations under which the field is not invariant are associated with null mass fields. In our case this particular group of transformations is associated to the quotient space $O(N)^3/O(N-1)^3$, which has dimension $3(N-1)$. If $\phi_{abc}=u_av_bw_c$, we can then see from  Eq.~\eqref{phivar} that Goldstone's bosons are indeed of the form $u^{i\bot}_av_bw_c$, $u_av^{i\bot}_bw_c$ and $u_av_bw^{i\bot}_c$ with $i=1,\dots,N-1$.
\subsection{$O(N-1)\times O(N)$-invariant solutions}

Another possible solution, considered by Diaz and Rosabal in Ref.~\cite{Diaz:2018eik}, is
\be \label{eq:vector-sol-1}
\phi_{abc} = u_a \d_{bc} + v_b \d_{ac} + w_c \d_{ab} \,,
\ee
which again breaks color symmetry unless $u=v=w$.\\
 Let us consider for simplicity the case $v=w=0$, whose stabilizer is $O(N-1)\times O(N)$ (or $O(N-1)\times {\rm Diag}[O(N)\times O(N)]$).
The field equations then become:
\be
0 = \l_2 + \left(\f{\l_d}{ N^2} +\l_p \left(\f{\a_1}{N}+ \f{\a_2+\a_3}{N^2}\right) +  \f{\l_t}{N^{3/2}} \right) u^2 \,,
\ee
and we see that again the scaling with $N$ is not homogeneous, and for $u^2\sim N$ only $\l_p$ contributes to the large-$N$ limit.
The solution is given by:
\be
u^2=\f{-\l_2N^2}{\l_d+\l_p(N\a_1+\a_2+\a_3)+\l_tN^{1/2}}.
\ee
Notice that the action evaluated on-shell scales like $N^{2}$, and it is negative for $\l_2<0$.

For this solution, the Hessian is
\be
\begin{split}
\cH_{abca'b'c'}&=\l_2\d_{aa'}\d_{bb'}\d_{cc'}+\f{\l_d}{N^3}(2u_au_{a'}\d_{bc}\d_{b'c'}+u^2N\d_{aa'}\d_{bb'}\d_{cc'})\\&
+\f{\l_p}{N^2}[\a_1(u^2\d_{aa'}\d_{bc}\d_{b'c'}+u_au_{a'}\d_{bc}\d_{b'c'}+Nu_au_{a'}\d_{bb'}\d_{cc'})\\&
+\a_2(u_au_{a'}\d_{b''c}\d_{b''c'}\d_{bb'}+u_au_{a'}\d_{b'c}\d_{bc'}+u^2\d_{b'c''}\d_{bc''}\d_{aa'}\d_{cc'})\\&
+\a_3(u_au_{a'}\d_{bc''}\d_{b'c''}\d_{cc'}+u_au_{a'}\d_{bc'}\d_{b'c}+u^2\d_{b''c}\d_{b''c'}\d_{aa'}\d_{bb'})]\\&
+\f{\l_t}{N^{3/2}}(u^2\d_{aa'}\d_{b'c}\d_{bc'}+u_au_{a'}\d_{b''c}\d_{b''c'}\d_{bb'}+u_au_{a'}\d_{bc''}\d_{b'c''}\d_{cc'}).
\end{split}
\label{eq:hess3}
\ee
We thus have to find the eigenvectors and the corresponding eigenvalues of this case. Let us suppose that a possible eigenvector could be written in the following way:
\be
\phi_{abc}=V_a M_{bc},
\ee
 where $M_{ab}$ is a $N\times N$ matrix whose form we have to determine. To do that, we can decompose $M_{ab}$ in its antisymmetric, symmetric traceless and trace components:
\be
M_{bc}=M_{bc}^{Tr}+M_{bc}^{S}+M_{bc}^{A}\,.
\ee

By rewriting the Hessian in a proper way, we can easily show that the complete basis of eigenvectors is given by:
\be
\begin{split}
	uM^{Tr},\,\,uM^{S},\,\,uM^{A},\,\,u^{\bot}M^{Tr},\,\,u^{\bot}M^{S},\,\,u^{\bot}M^{A}\,.
	\label{eig2}
\end{split}
\ee 
In order to do that, let us define:
\be
\begin{split}
	&A_{bcb'c'}=\f{1}{2}(\d_{bb'}\d_{cc'}-\d_{bc'}\d_{b'c}) \,,\\&
	S_{bcb'c'}=\f{1}{2}(\d_{bb'}\d_{cc'}+\d_{bc'}\d_{b'c}-\f{2}{N}\d_{bc}\d_{b'c'}) \,,\\&
	T_{bcb'c'}=\f{1}{N}\d_{bc}\d_{b'c'} \,.
\end{split}
\ee
These are the projectors on the antisymmetric, symmetric traceless and trace subspaces.
They satisfy the following identities:
\be
\begin{split}
&\d_{bb'}\d_{cc'}=S_{bcb'c'}+T_{bcb'c'}+A_{bcb'c'}\\&
\d_{bc'}\d_{b'c}=S_{bcb'c'}+T_{bcb'c'}-A_{bcb'c'}\\&
\d_{bc}\d_{b'c'}=N\,T_{bcb'c'}.
\end{split}
\ee
By looking at the form of Eq.~\eqref{eq:hess3}, we can then rewrite:
\be
\begin{split}
\cH_{abca'b'c'}&=\f{\a_1\l_2N\l_p+2\l_2\sqrt{N}\l_t}{\l_d+\l_p (\a_2+\a_3+\a_1N)+\sqrt{N} \l_t}\d_{aa'}A_{bcb'c'}\\&
+\f{\a_1\l_2N\l_p}{\l_d+\l_p(\a_2+\a_3+\a_1N)+\sqrt{N}\l_t}\d_{aa'}S_{bcb'c'}\\&
+(\f{2 \l_t}{N^{3/2}}+\f{\a_1 \l_p}{N})u_au_{a'}A_{bcb'c'}\\&
+\left[\f{(2 (\a_2+\a_3)+\a_1N)\l_p+2\sqrt{N}\l_t}{N^2}\right]u_au_{a'}S_{bcb'c'}\\&
+2\left[\f{\l_d+(\a_2+\a_3+\a_1N)\l_p+\sqrt{N}\l_t}{N^3}\right]u_au_{a'}T_{bcb'c'}.
\end{split}
\ee
The eigenvalues of each eigenvector from Eq.~\eqref{eig2} are then:
\be
\begin{split}
	&\cH_{abca'b'c'}(u_{a'}M^A_{b'c'})=0;\\&
	\cH_{abca'b'c'}(u_{a'}M^S_{b'c'})=-\f{2 \l_2 \left(\sqrt{N} \l_t+(\a_2+\a_3)\l_p\right)}{\l_d+\l_p (\a_2+\a_3+\a_1N)+\sqrt{N} \l_t}u_aM^S_{bc};\\&
	\cH_{abca'b'c'}(u_{a'}M^{Tr}_{b'c'})=-\f{2\l_2}{N}u_aM^{Tr}_{bc};\\&
	\cH_{abca'b'c'}(u^{\bot}_{a'}M^A_{b'c'})=\f{\a_1 \l_2N\l_p+2\l_2 \sqrt{N} \l_t}{\l_d+\l_p (\a_2+\a_3+\a_1 N)+\sqrt{N} \l_t}u^{\bot}_{a}M^A_{bc};\\&
	\cH_{abca'b'c'}(u^{\bot}_{a'}M^S_{b'c'})=\f{\a_1\l_2 N \l_p}{\l_d+\l_p (\a_2+\a_3+\a_1N)+\sqrt{N}\l_t}u^{\bot}_{a}M^S_{bc};\\&
	\cH_{abca'b'c'}(u^{\bot}_{a'}M^{Tr}_{b'c'})=0.
\end{split}
\ee
The Goldstone's modes are $uM^{A}$ and $u^{\bot}M^{Tr}$.
Furthermore, we can again observe the presence of negative eigenvalues that indicate the instability of the solution.\\
\subsection{$O(N^2-N)$-invariant solutions at  $\a_2=\a_3=\l_t=0$}

The model with $\a_2=\a_3=\l_t=0$ has enhanced symmetry $O(N)\times O(N^2)$. It is straightforward to find a symmetry breaking solution: it suffices to find $N$ orthonormal vectors in $\mathbb{R}^{N^2}$. In fact, let $\phi_{abc} = v_{aA}$, with for example $A=b+N(c-1)=1\ldots N^2$, such that
\be
v_{aA} v_{a' A} = N^2 \d_{aa'} v^2\,.
\ee
Then, the equations of motion reduce to
\be
0=\l_2 + (\l_d + \l_p \a_1) v^2 \,,
\ee
which have a real solution for $(\l_d + \l_p \a_1)/\l_2<0$.
Any given solution of this type is clearly invariant under the group of orthogonal transformations in the space orthogonal to the $N$ vectors, i.e.\ $O(N^2-N)$. Furthermore, it is also invariant under simultaneous $O(N)$ rotations acting on the vector components and on the vector labels. In other words, the stability group is $O(N)\times O(N^2-N)$.

For this solution, the Hessian has the form:
\be
\cH_{aAa'A'}=\f{2\l_d}{N^3}v_{aA}v_{a'A'}+\f{\l_p\a_1}{N^2}v_{a'A}v_{aA'}+\f{\l_p\a_1}{N^2}\d_{aa'}v_{a''A}v_{a''A'}\,.
\ee
Given the pattern of symmetry breaking, the number of Goldstone's bosons is given by the number of generators of $O(N^2)$ minus the number of generators of $O(N^2-N)$, i.e.:
\be
\f{N^2 (N^2-1)}{2} - \f{(N^2-N) (N^2-N-1)}{2} 
=N^3-\f{N(N+1)}{2} \,.
\ee
As a basis for the Goldstone's bosons we can choose the tensors $r^I_{aA}$ defined by:
\be
r^I_{aA}=L^I_{AA'}v_{aA'}\,,\;\;\;\;\; 
\; I=1,\dots,N^3-\f{N(N+1)}{2} \,,
\ee 
where $L^I$ are the generators of orthogonal transformations in the space orthogonal to the $N$ vectors, i.e.\ $O(N^2-N)$.
Regarding the other eigenvectors, one can show that the tensors, together with $v_{aA}$, defined by:
\be
u_{aA}\equiv S^{\a}_{aa'}v_{a'A}\,\, \a=1,\dots,\f{N(N+1)}{2}-1\,,
\ee
 where $S^\a$ are symmetric traceless matrices, are eigenvectors of the Hessian with eigenvalues:
\be
\begin{split}
&\cH_{aAa'A'}v_{a'A'}=-2\l_2v_{aA};\\&
\cH_{aAa'A'}u_{a'A'}=-2\l_2\,\f{\l_p\a_1}{\l_p\a_1+\l_d}u_{aA}\,.
\end{split}
\ee
Therefore, for $\l_2<0$ the solution is stable.


\addcontentsline{toc}{section}{References}

\providecommand{\href}[2]{#2}\begingroup\raggedright\endgroup


\end{document}